\pdfoutput=1

\documentclass[11pt,twoside,a4paper,cmspaper,final,collab]{cms-tdr}
\def\svnVersion{89854}\def\cmsCernNo{2011-159}\def\cmsCernDate{\today}\def\cmsMessage{Submitted to Physical Review D}

\begin{document}\cmsNoteHeader{EWK-11-003}

\hyphenation{had-ron-i-za-tion}
\hyphenation{cal-or-i-me-ter}
\hyphenation{de-vices}

\RCS$Revision: 81590 $
\RCS$HeadURL: svn+ssh://alverson@svn.cern.ch/reps/tdr2/papers/EWK-11-003/trunk/EWK-11-003.tex $
\RCS$Id: EWK-11-003.tex 81590 2011-10-19 20:36:39Z gritsan $
\newcommand{\sss}{\scriptscriptstyle}
\newlength{\cmsfigwid}
\ifthenelse{\boolean{cms@external}}{\setlength\cmsfigwid{0.85\columnwidth}}{\setlength\cmsfigwid{0.6\textwidth}}
\newlength{\cmsSmallfigwid}
\ifthenelse{\boolean{cms@external}}{\setlength\cmsSmallfigwid{0.55\columnwidth}}{\setlength\cmsSmallfigwid{0.4\textwidth}}
\ifthenelse{\boolean{cms@external}}{\newcommand{\xleft}{top\xspace}}{\newcommand{\xleft}{left\xspace}}
\ifthenelse{\boolean{cms@external}}{\newcommand{\xright}{bottom\xspace}}{\newcommand{\xright}{right\xspace}}
\ifthenelse{\boolean{cms@external}}{\newcommand{\xLeft}{Top\xspace}}{\newcommand{\xLeft}{Left\xspace}}
\ifthenelse{\boolean{cms@external}}{\newcommand{\xRight}{Bottom\xspace}}{\newcommand{\xRight}{Right\xspace}}
\cmsNoteHeader{EWK-11-003} % This is over-written in the CMS environment: useful as preprint no. for export versions
\title{Measurement of the weak mixing angle with the Drell--Yan process in proton-proton collisions at the LHC}

\date{\today}

\abstract{
A multivariate likelihood method to measure electroweak couplings with the Drell--Yan
process at the LHC is presented. The process is described by the dilepton rapidity,
invariant mass, and decay angle distributions.
The decay angle ambiguity due to the unknown assignment of the scattered constituent
quark and antiquark to the two protons in a collision is resolved statistically using correlations
between the observables.
The method is applied to a sample of dimuon events
from proton-proton collisions at $\sqrt{s} = 7\,$TeV collected by
the CMS experiment at the LHC, corresponding to an integrated luminosity of 1.1~fb$^{-1}$.
From the dominant $\cPqu\cPaqu, \cPqd\cPaqd\to \gamma^*/\cPZ\to \mu^-\mu^+$ process, the effective
weak mixing angle parameter is measured to be
$\sin^2\theta_\text{eff}=0.2287 \pm 0.0020~(\text{stat.}) \pm 0.0025~(\text{syst.})$\,.
This result is consistent with measurements from other processes, as expected within
the standard model.
}

\hypersetup{%
pdfauthor={CMS Collaboration},%
pdftitle={Measurement of the weak mixing angle with the Drell-Yan process in proton-proton collisions at the LHC},%
pdfsubject={CMS},%
pdfkeywords={CMS, physics, Drell-Yan}}

\maketitle %maketitle comes after all the front information has been supplied

\section{Introduction}
\label{sec:intro}

\begin{figure}[htbp]
\begin{center}
\includegraphics[width=\cmsSmallfigwid]{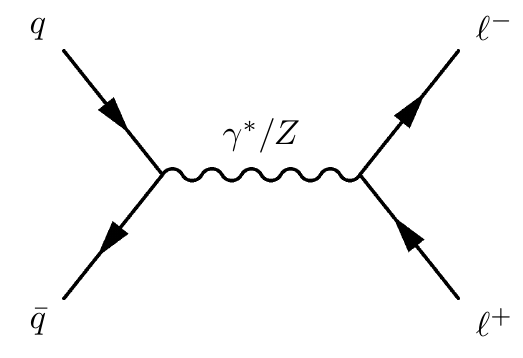}
~~~~~~~~~
\includegraphics[width=\cmsSmallfigwid]{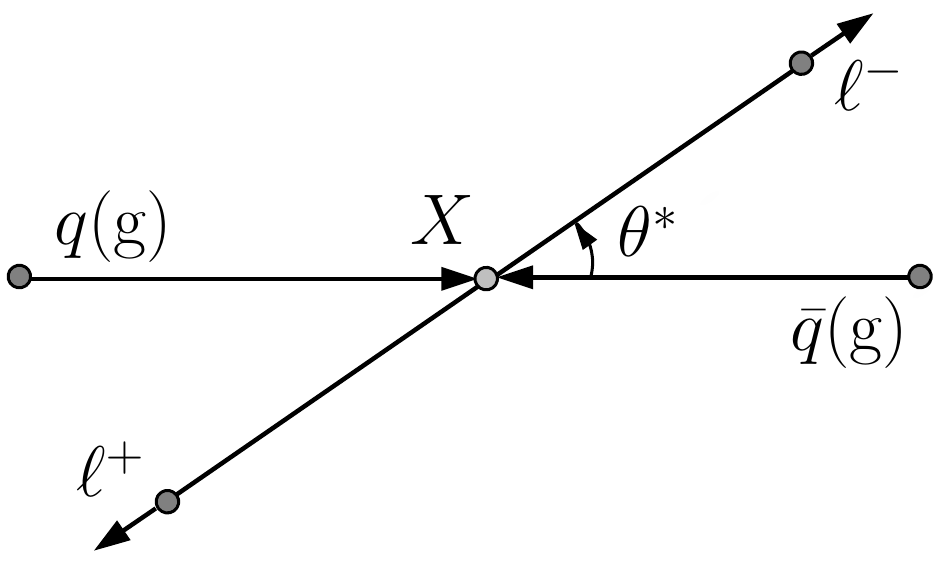}
\end{center}
\caption{
\xLeft: diagram describing the SM process
$q\overline{q}\rightarrow \gamma^{*}/\cPZ \rightarrow\ell^-\ell^+$.
\xRight: definition of the angle $\theta^*$ in the production and decay of an intermediate state $X$,
such as \cPg\cPg\ or $q\overline{q}\rightarrow X \rightarrow\ell^-\ell^+$.
\label{fig:angles}
}
\end{figure}

The main goal of the Large Hadron Collider (LHC)~\cite{lhc-paper} is to explore physics at the
TeV energy scale via proton-proton collisions.
The Drell--Yan process~\cite{Drell:1970wh} occurs through the annihilation
of a quark from one proton with an antiquark from the other proton,
creating a virtual neutral gauge boson ($\gamma^*$ or $\cPZ$)
that subsequently decays to a pair of oppositely charged leptons, as shown in Fig.~\ref{fig:angles}.
This annihilation process could reveal the existence of a new neutral gauge
boson~\cite{London:1986,Rosner:1987a} or uncover deviations from the
standard model of particle physics (SM) in elementary fermion couplings
to the known neutral electroweak bosons.
Besides quark-antiquark annihilation, new resonances decaying into lepton pairs
may also be produced at  the LHC via the gluon-fusion mechanism.

Analysis of electron-positron annihilations into dilepton or $\cPqb\cPaqb$ pairs at LEP and
SLC~\cite{Schael:2005ema} led to high-precision measurements of the electroweak $\cPZ$-boson
couplings to fermions. The measurement of the weak mixing angle parameter $\sin^2\theta_{\PW}$
was performed to a precision of $\sim0.1\%$. In the SM,
the weak mixing angle $\theta_\PW$ describes the rotation of the original $\PW^0$ and $\cmsSymbolFace{B}^0$
vector boson states into the observed $\gamma$ or $\cPZ$ bosons as a result of spontaneous
symmetry breaking~\cite{Weinberg:1967tq}. Within the SM,
$\sin^2\theta_{\PW}$ is the only free parameter that fixes the relative
couplings of all fermions to the $\gamma$ or $\cPZ$.

Measurements of the weak mixing angle with different initial and final fermion-antifermion
states $f_1\overline{f_1}\rightarrow \gamma^{*}/\cPZ \rightarrow f_2\overline{f_2}$
tests the universality of the fermion/gauge-boson interactions and predictions of the SM.
For example, LEP measurements~\cite{Schael:2005ema}
of the inclusive hadronic charge asymmetry provided a measurement
of the couplings of light quarks compared to leptons and $\cPqb$ quarks.
The NuTeV collaboration measured $\sin^2\theta_\PW$ to precision of about $1\%$
from neutrino and antineutrino deep inelastic scattering on nucleons~\cite{Zeller:2001hh}.
The CDF and D0~experiments~\cite{CDF:2005, D0:2008, Abazov:2011ws} at the Tevatron
reached a similar precision with the Drell--Yan process in proton-antiproton collisions,
as did the H1 experiment~\cite{Aktas:2005iv} with electron-proton scattering at HERA.
In this paper, we measure the $\sin^2\theta_{\PW}$ parameter to a precision $\sim1\%$
in the proton-proton Drell--Yan process at the LHC with the CMS experiment.

The proton-proton collisions at the LHC pose new challenges compared to previous collider
experiments. The interference of the axial-vector and vector couplings leads to an asymmetry
in the distribution of the polar angle of the lepton with respect to the direction of the
constituent quark from the incoming proton in the quark-antiquark annihilation.
This type of ``forward-backward" asymmetry has been the primary
measurement used to extract the couplings and $\sin^2\theta_{\PW}$ at the LEP, SLC, and
Tevatron experiments. However, because of the symmetric proton-proton collision at the
LHC, the direction of the quark is not known and can be deduced only on a statistical
basis~\cite{Rosner:1987a, Fisher:1994pw, Dittmar:1997, Altarelli:2000ye} 
using the boost direction of the dilepton
final state, because of the higher probability for a valence quark from one of
the incoming protons to provide the boost.
We have developed methods that allow a per-event analytical likelihood description to extract the
maximal information about the process at the LHC. This technique exploits more information
than the conventional forward-backward asymmetry approach, and the distribution
of the polar angle as a function of both dilepton rapidity and mass is an essential
component of this method. The technique has applications both for high-precision
electroweak measurements and for rare-process searches at the LHC.

In this paper, we present a multivariate analysis that uses the full information about the Drell--Yan
process $q\overline{q}\rightarrow \gamma^{*}/\cPZ \rightarrow\ell^-\ell^+$, parameterized as a function
of the dilepton rapidity $Y$, invariant mass squared $\hat{s}$, and decay angle $\theta^{*}$.
This process offers a relatively simple environment for the development of the matrix-element
analysis techniques for resonance polarization studies at the LHC. Encouraged by feasibility
studies of an analytical matrix-element approach in Ref.~\cite{Gao:2010qx}, we use a formalism
based on an analytical description of the elementary interaction (Section~\ref{sec:pheno}).
The method is applied to a sample of proton-proton collisions at a center-of-mass
energy $\sqrt{s}$ = 7\TeV,  corresponding to an integrated luminosity of
1.1\fbinv recorded by the CMS experiment (Section~\ref{sec:reco}).
We include a description of detector effects in the analytical likelihood model
(Section~\ref{sec:method}) and pay particular attention to systematic effects
(Section~\ref{sec:systematics}). The result is consistent with measurements
in other processes, as expected within the SM (Section~\ref{sec:results}).

\section{Phenomenology of the Drell--Yan Process at the LHC}
\label{sec:pheno}

The philosophy of the multivariate likelihood analysis is to first produce a phenomenological
model of the process and then introduce detector effects into the model.
The parameters of the model may either be fixed to the best known values or
left free in the fit, to be determined from data. These parameters may include the
physical quantities of interest, such as $\sin^2\theta_{\PW}$, or a description of
detector effects, such as a correction for the momentum scale in the track reconstruction.
Therefore, we start with a discussion of a phenomenological model, and then proceed
to detector-specific effects in the application of the analysis to CMS data.

The tree-level coupling of a spin-one gauge boson to fermions is described by
\begin{eqnarray}
\epsilon^\mu \overline{u}_{f} \gamma_\mu  e \left(  \rho_{\sss V} - \rho_{\sss A}\gamma_5 \right) v_{f},
\label{eq:ampl-spin1-qq}
\end{eqnarray}
where $v_f$ and $\overline{u}_f$ are the Dirac spinors of the fermion ($f$) and antifermion ($\overline{f}$),
$\epsilon^\mu$ is the polarization vector of the spin-one boson, and
$\rho_{\sss V}$ and $\rho_{\sss A}$ are the vector and axial-vector couplings.
The couplings of the SM gauge bosons $\gamma$ and $\cPZ$ are
given in Table~\ref{table-vectorcurrent}.
In the limit of negligible fermion masses, which is a good approximation for
both quarks and leptons in the Drell--Yan process near the $\cPZ$-boson mass,
only two helicity states of the fermions are possible.  They correspond to amplitudes
$A_{\uparrow\downarrow}\propto(\rho_{\sss V}-\rho_{\sss A})$
and $A_{\downarrow\uparrow}\propto-(\rho_{\sss V}+\rho_{\sss A})$.
\begin{table}[h!]
\begin{center}
\caption{
Vector and axial-vector couplings of the SM gauge bosons to the charged fermion fields~\cite{Nakamura:2010zzi}.
}
\begin{tabular}{lcc}
\hline\hline
\vspace{0.1cm}
  & $\rho_{\sss V}$ & $\rho_{\sss A}$  \\
\hline
\vspace{-0.3cm} &  &   \\
\vspace{0.1cm}
 $\gamma\to \Pem\Pep, \Pgmm\Pgmp, \Pgt^-\Pgt^+$ & $-1$ & 0  \\
\vspace{0.1cm}
 $\gamma\to \cPqu\cPaqu, \cPqc\cPaqc, \cPqt\cPaqt$ & $+2/3$ & 0  \\
\vspace{0.1cm}
 $\gamma\to \cPqd\cPaqd, \cPqs\cPaqs, \cPqb\cPaqb$ & $-1/3$ & 0  \\
\vspace{0.1cm}
 $\cPZ\to \Pem\Pep, \Pgmm\Pgmp, \Pgt^-\Pgt^+$   & $\frac{-3+12\sin^2\theta_\PW}{6\sin(2\theta_\PW)}$
                 & $\frac{-1}{2\sin(2\theta_\PW)}$  \\
\vspace{0.1cm}
 $\cPZ\to \cPqu\cPaqu, \cPqc\cPaqc, \cPqt\cPaqt$ & $\frac{+3-8\sin^2\theta_\PW}{6\sin(2\theta_\PW)}$
                 & $\frac{+1}{2\sin(2\theta_\PW)}$  \\
\vspace{0.1cm}
 $\cPZ\to \cPqd\cPaqd, \cPqs\cPaqs, \cPqb\cPaqb$ & $\frac{-3+4\sin^2\theta_\PW}{6\sin(2\theta_\PW)}$
                 & $\frac{-1}{2\sin(2\theta_\PW)}$   \\
\hline\hline
\end{tabular}
\label{table-vectorcurrent}
\end{center}
\end{table}

The parton-level cross section for the Drell--Yan process can be
expressed with the help of the Wigner $d^J_{m,m^\prime}$ matrix, assuming the
spin $J=1$ intermediate states $\gamma^*$, $\cPZ$, and possible new unknown
contributions (indicated by an ellipsis below), as
\begin{widetext}
\begin{eqnarray}
\hat\sigma_{q\overline{q}}(\hat{s}, \cos\theta^*; \theta_{\PW})
~~~~\propto~~~~ \frac{1}{\hat{s}}
\sum_{\chi_1,\chi_2,\lambda_1,\lambda_2=\uparrow,\downarrow}
\left (2J+1\right)\Bigl(d^{J=1}_{\chi_1-\chi_2,\lambda_1-\lambda_2}(\theta^*)\Bigr)^2
~~~~~~~~~~~~~~~~~~~~~~~~~
\nonumber\\
\times
\Biggl| A^{q\overline{q}\to\gamma}_{\chi_1,\chi_2} A^{\gamma\to\ell\ell}_{\lambda_1,\lambda_2}
 + A^{q\overline{q}\to \cPZ}_{\chi_1,\chi_2}({\theta_\PW}) A^{\cPZ\to\ell\ell}_{\lambda_1,\lambda_2}({\theta_\PW})
 \times\frac{\hat{s}}{ (\hat{s}-{m}_\cPZ^2)+i{m}_\cPZ\Gamma_\cPZ} + \ldots \Biggr|^2
 \,,
\label{eq:production-general}
\end{eqnarray}
\end{widetext}
where $\chi_1,\chi_2,\lambda_1,$ and $\lambda_2$ are the helicity states of the
quark, antiquark, lepton, and antilepton, and the dilepton decay angle $\theta^*$
is defined in the center-of-mass frame of the dilepton system as shown in Fig.~\ref{fig:angles}.
The effects of transverse motion of the incoming constituent quark and antiquark
in their annihilation are minimized by using the Collins--Soper frame~\cite{Collins:1977}.
In this frame, the angle $\theta^{*}$
is defined as the angle between the lepton momentum and a $z^\prime$-axis that bisects
the angle between the direction of one proton and the direction opposite that of
the other proton in the dilepton rest frame. Should there be new states contributing
to the cross section with the same or different spin,
they will either contribute with interference if produced in
$q\overline{q}$ annihilation, or without interference if produced in gluon fusion (\cPg\cPg) for example.

The parton-level cross section in Eq.~(\ref{eq:production-general}) can be further
simplified for the SM intermediate states $\gamma^*$ and $\cPZ$ with spin $J=1$ as
\begin{widetext}
\begin{eqnarray}
\hat{\sigma}_{q\overline{q}}(\hat{s},\cos\theta^*;\theta_{\PW}) ~~~~\propto ~~~~
\frac{3\,\left(\rho_{\sss V}^{q\overline{q}\to\gamma}\right)^2\left(\rho_{\sss V}^{\gamma\to\ell\ell}\right)^2 }{2\,\hat{s}} \times (1+\cos^2\theta^\ast)
~~~~~~~~~~~~~~~~~~~~~~~~~~~~~~~~~~~~~~~~~~~~~~~~~~~
 \nonumber \\
+ \frac{3}{2} \frac{\hat{s}}{(\hat{s}-m_\cPZ^2)^2+m_\cPZ^2\Gamma_\cPZ^2}
\times\biggl[
\left( \left(\rho_{\sss V}^{q\overline{q}\to \cPZ}\right)^2 + \left(\rho_{\sss A}^{q\overline{q}\to \cPZ}\right)^2\,\right)
\left( \left(\rho_{\sss V}^{Z\to\ell\ell}\right)^2 + \left(\rho_{\sss A}^{\cPZ\to\ell\ell}\right)^2\,\right)
\left(1+\cos^2\theta^\ast\right)
 \nonumber \\
~~~~~~~~~~~~~~~~~~~~~~~~~~~~~~~~~~
+ 8\,\rho_{\sss V}^{q\overline{q}\to\cPZ}\rho_{\sss A}^{q\overline{q}\to \cPZ} \rho_{\sss V}^{Z\to\ell\ell}\rho_{\sss A}^{\cPZ\to\ell\ell} \cos\theta^\ast
\biggr] \nonumber \\
+ \frac{3(\hat{s}-m_\cPZ^2)  \rho_{\sss V}^{q\overline{q}\to\gamma}\rho_{\sss V}^{\gamma\to\ell\ell}}{(\hat{s}-m_\cPZ^2)^2+m_\cPZ^2\Gamma_\cPZ^2}
\times\biggl[
\rho_{\sss V}^{q\overline{q}\to \cPZ}\rho_{\sss V}^{\cPZ\to\ell\ell}\left(1+\cos^2\theta^\ast\right)
+2\,\rho_{\sss A}^{q\overline{q}\to \cPZ}\rho_{\sss A}^{\cPZ\to\ell\ell}\cos\theta^\ast
\biggr]\,.
\label{eq:mass-alltogether}
\end{eqnarray}
\end{widetext}

Equation~(\ref{eq:mass-alltogether}) implies that the asymmetry in the polar angle
arises from terms linear in $\cos\theta^*$.
The value of the asymmetry depends on the dilepton invariant mass $m=\sqrt{\hat{s}}$,
quark flavor $\cPq$, and weak mixing angle $\theta_\PW$. It could also be affected by
deviations of couplings from SM expectations or by the presence of new contributions.

The differential cross section of the proton-proton Drell--Yan process can be expressed
as a product of the parton-level cross section in
Eqs.~(\ref{eq:production-general}) or~(\ref{eq:mass-alltogether})
and the parton distribution functions (PDFs) $ f_a(x_i,\hat{s})$~\cite{DeRoeck:2011na}
describing the the probability for partons of type $a$ to have a fraction $x_i$ of the proton
momentum $p=\sqrt{s}/2$:
\begin{widetext}
\begin{eqnarray}
\frac{{d\sigma_{\Pp\Pp}}(px_1, px_2, \cos\theta^\ast;\theta_{\PW})}
{dx_1 \, dx_2 \, d\cos\theta^\ast}
& \propto &
\!\!\!\!\!\!\! \sum_{q=\cPqu,\cPqd,\cPqs,\cPqc,\cPqb}
\bigg(
{\hat\sigma_{q\overline{q}}}(\hat{s}, \text{sgn}(x_1-x_2)\cos\theta^\ast;\theta_{\PW}) \,
 f_q(x_1,\hat{s}) \;  f_{\bar{q}} (x_2,\hat{s}) \nonumber\\
& &
+ ~ {\hat\sigma_{q\overline{q}}}(\hat{s}, \text{sgn}(x_2-x_1)\cos\theta^\ast;\theta_{\PW}) \,
 f_q(x_2,\hat{s}) \;  f_{\bar{q}} (x_1,\hat{s})
\bigg) \,.
\label{eq:dilution-amplitude-ppbar}
\end{eqnarray}
\end{widetext}
The expression $\text{sgn}(x_1-x_2)$ refers to the sign of the difference $(x_1-x_2)$,
reflecting the fact that the quark direction is assumed to coincide with the boost of the
$q\overline{q}$ system. This assumption introduces a dilution in the odd-power terms in $\cos\theta^*$.

It is convenient to convert from the two variables $(x_1,x_2)$ to $(Y, \hat{s})$,
the dilepton rapidity and the square of the dilepton mass, as
\begin{eqnarray}
\label{eq:rapidity}
Y  & = & \frac{1}{2}\ln\left(\frac{\hat{E}+\hat{p}_z}{\hat{E}-\hat{p}_z}\right)
~=~ \frac{1}{2}\ln\left(\frac{x_1}{x_2}\right) \,, \\
\label{eq:rapidity2}
\hat{s}  & = & \hat{E}^2-\hat{p}^2 ~=~ x_1\,x_2\,{s} \,,
\end{eqnarray}
where $\hat{E}$ is the dilepton system energy,  $\hat{p}$ and $\hat{p}_z$ are its momentum
and longitudinal momentum in the laboratory frame, and $\hat{p}=|\hat{p}_z|$ at
leading order in QCD.

After transformation of the variables, the Drell--Yan process in proton-proton
interactions can be expressed as follows:
\begin{widetext}
\begin{eqnarray}
&& \frac{{d\sigma_{pp}}(Y, \hat{s}, \cos\theta^\ast;\theta_{\PW})} { dY \, d\hat{s}  \, d\cos\theta^\ast} \propto
\sum_{q=\cPqu,\cPqd,\cPqs,\cPqc,\cPqb}
\Big[
\hat\sigma_{q\overline{q}}^\text{even}(\hat{s},\cos^2\theta^\ast;\theta_{\PW})
\nonumber \\
&&
~~~~~~~~~~~~~~~~~~~~~~~~~~~~~~~~~~~~~~~~~~~~~~
+D_{q\overline{q}}(\hat{s}, Y)\times
\hat\sigma_{q\overline{q}}^\text{odd}(\hat{s},\cos\theta^\ast;\theta_{\PW})
\Big]
\times F_{q\overline{q}}(\hat{s}, Y)
\,.
\label{eq:dilution-amplitude-3}
\end{eqnarray}
The parton factor is defined as
\begin{eqnarray}
F_{q\overline{q}}(\hat{s}, Y)=
{
 f_q\left(e^{+Y}\!\sqrt{ \hat{s}/{s}}, \,\hat{s} \right) \;  f_{\bar{q}}\left(e^{-Y}\!\sqrt{ \hat{s} /s}, \,\hat{s} \right)
+
 f_q\left(e^{-Y}\!\sqrt{ \hat{s}/{s}}, \,\hat{s} \right) \;  f_{\bar{q}}\left(e^{+Y}\!\sqrt{ \hat{s} /s}, \,\hat{s} \right)
}
\,,
\label{eq:dilution-amplitude-parton}
\end{eqnarray}
and the dilution factor is defined as
\begin{eqnarray}
\label{eq:dilution-amplitude-factor}
D_{q\overline{q}}(\hat{s}, Y) =
\frac
{
 f_q\left(e^{+|Y|}\sqrt{ \hat{s}/{s}}, \,\hat{s} \right) \;  f_{\bar{q}}\left(e^{-|Y|}\sqrt{ \hat{s}/s}, \,\hat{s} \right)
-
 f_q\left(e^{-|Y|}\sqrt{ \hat{s}/{s}}, \,\hat{s} \right) \;  f_{\bar{q}}\left(e^{+|Y|}\sqrt{ \hat{s}/s}, \,\hat{s} \right)
}
{
 f_q\left(e^{+Y}\!\sqrt{ \hat{s}/{s}}, \,\hat{s} \right) \;  f_{\bar{q}}\left(e^{-Y}\!\sqrt{ \hat{s} /s}, \,\hat{s} \right)
+
 f_q\left(e^{-Y}\!\sqrt{ \hat{s}/{s}}, \,\hat{s} \right) \;  f_{\bar{q}}\left(e^{+Y}\!\sqrt{ \hat{s} /s}, \,\hat{s} \right)
}
\,.
\end{eqnarray}

The two components of the parton cross section
contain either an even or odd power of $\cos\theta^*$
\begin{eqnarray}
&&\hat\sigma_{q\overline{q}}^\text{even}(\hat{s},\cos^2\theta^\ast;\theta_{\PW})=
\frac{1}{2}\left( {\hat\sigma_{q\overline{q}}}( \hat{s}, +\cos\theta^\ast;\theta_{\PW}) + {\hat\sigma_{q\overline{q}}}( \hat{s}, -\cos\theta^\ast;\theta_{\PW}) \right)
\label{eq:cs_even}
\,,\\
&&\hat\sigma_{q\overline{q}}^\text{odd}(\hat{s},\cos\theta^\ast;\theta_{\PW})=
\frac{1}{2}\left( {\hat\sigma_{q\overline{q}}}( \hat{s}, +\cos\theta^\ast;\theta_{\PW}) - {\hat\sigma_{q\overline{q}}}( \hat{s}, -\cos\theta^\ast;\theta_{\PW}) \right)
\,.
\label{eq:cs_odd}
\end{eqnarray}

\end{widetext}

\begin{figure}[htbp]
\begin{center}
\includegraphics[width=\cmsfigwid]{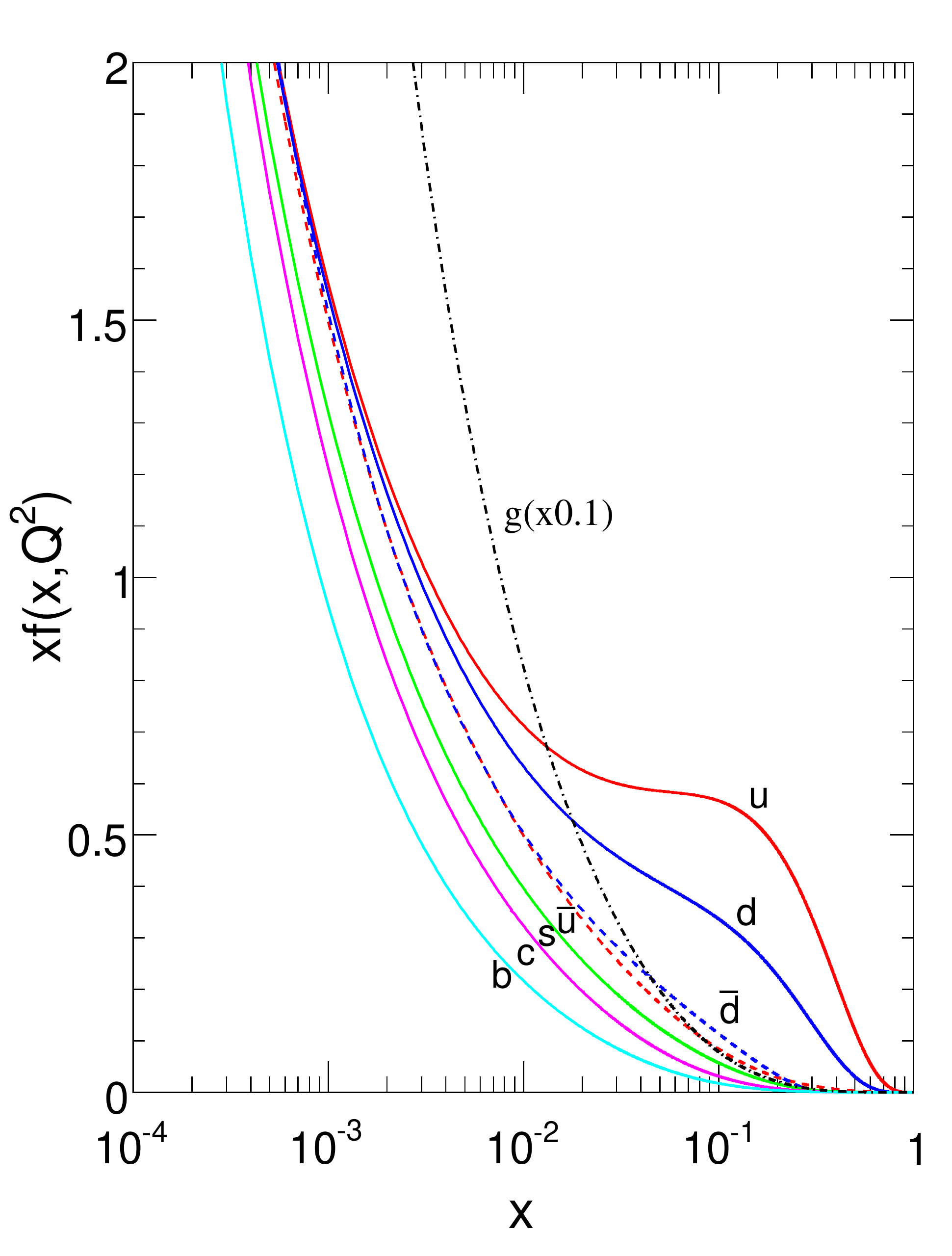}
\end{center}
\caption{
Analytical parameterization of the parton distribution functions $x f_a(x,Q^2)$ at $Q=100\GeV$
using the \textsc{cteq6}~\cite{CTEQ6:2004} numerical computation for the various quarks, antiquarks,
and the gluon. The gluon distribution is scaled by a factor of 0.1.
\label{fig:pdfs}
}
\end{figure}

\begin{figure}[htbp]
\begin{center}
\includegraphics[width=\cmsfigwid]{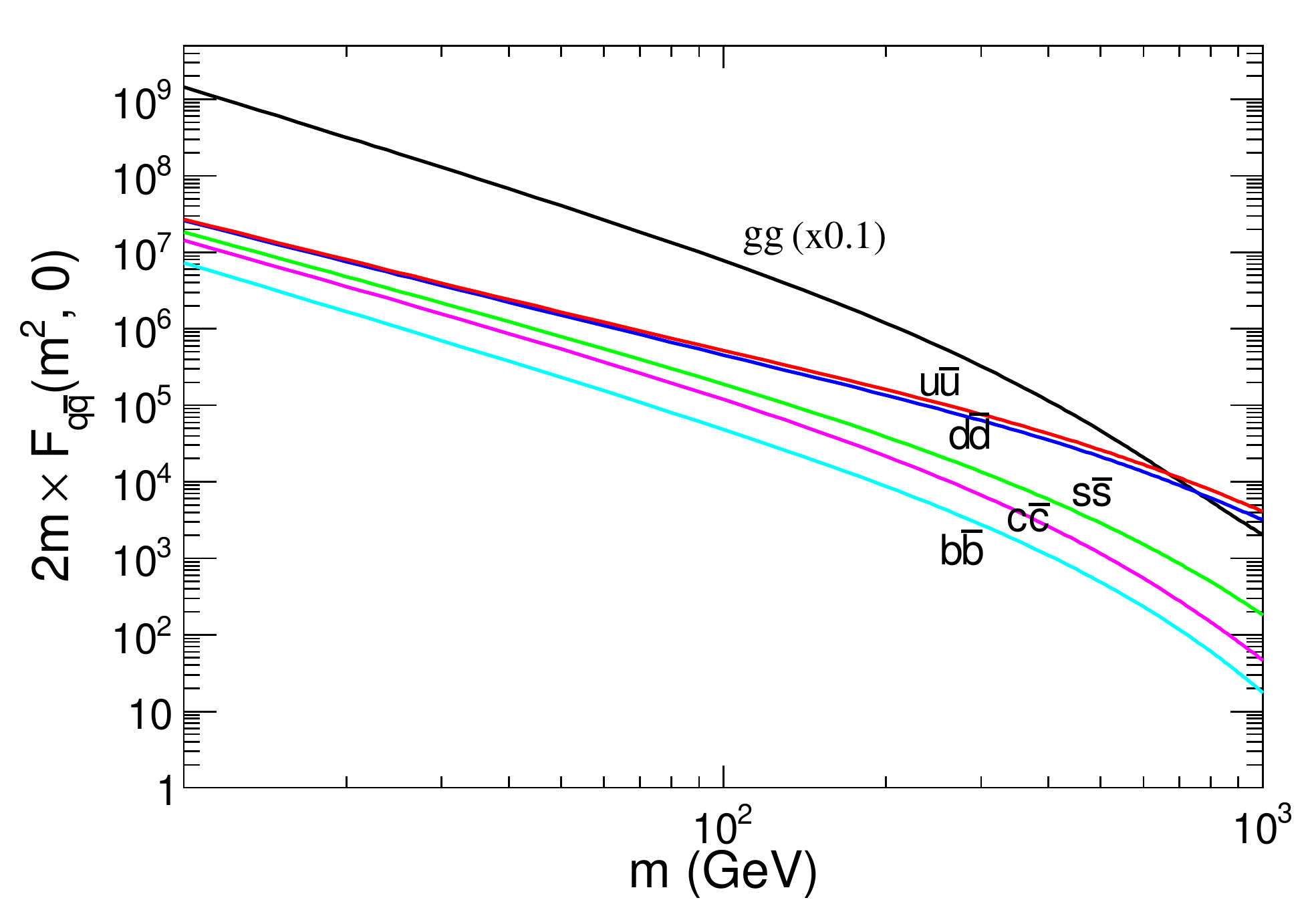}
\end{center}
\caption{
Distribution of the effective cross section factor $2mF_{q\overline{q}}(m^2,Y=0)$
defined in Eq.~(\ref{eq:dilution-amplitude-parton}) for five quark flavors
(from top to bottom $q=\cPqu,\cPqd,\cPqs,\cPqc,\cPqb$) for proton-proton collision energies of 7\TeV
as a function of the dilepton mass $m$. An equivalent factor for gluon-fusion production
is shown for comparison and is scaled by a factor of 0.1.
\label{fig:factor}
}
\end{figure}

\begin{figure}[htbp]
\begin{center}
\includegraphics[width=\cmsfigwid]{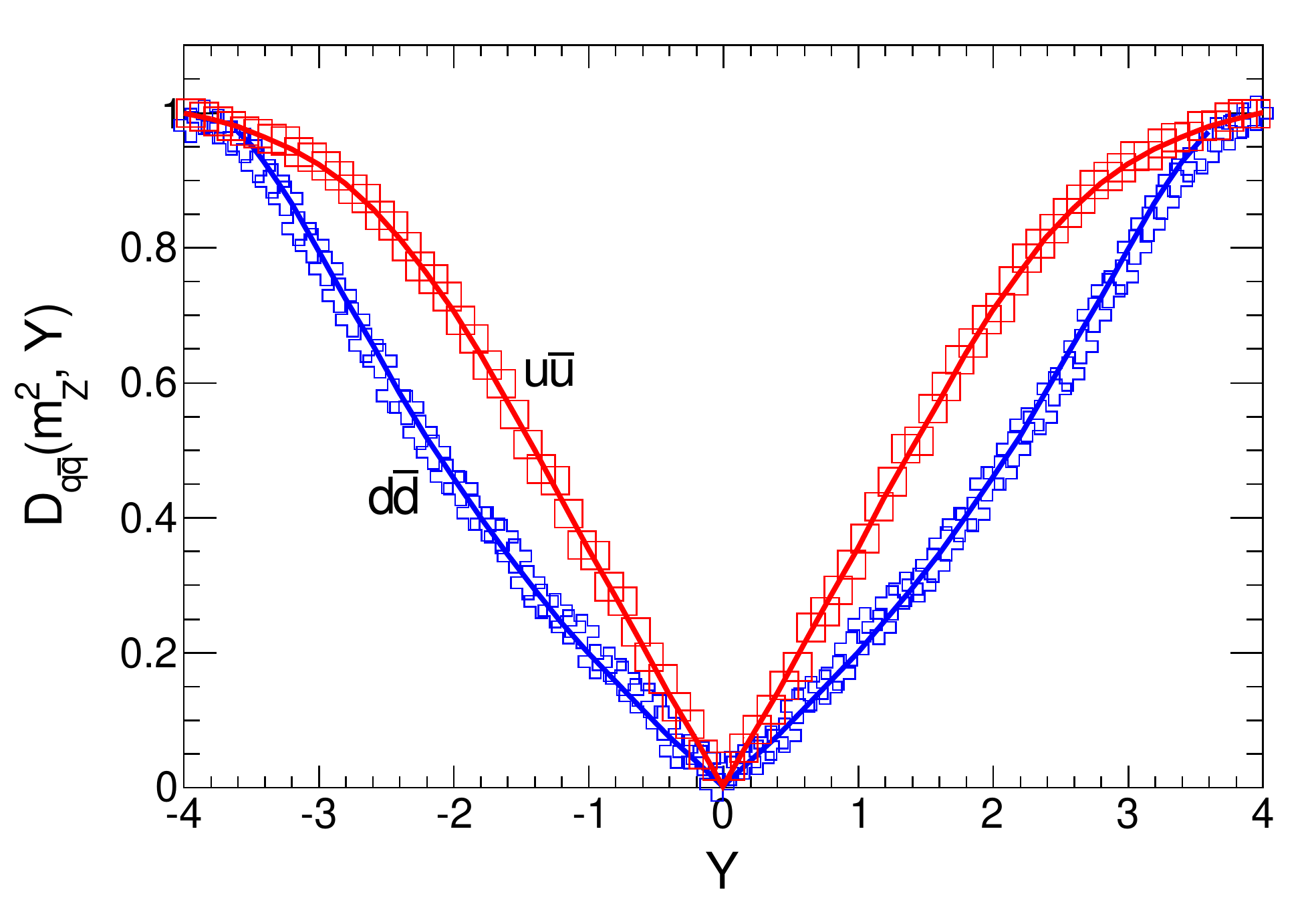}
\end{center}
\caption{
The dilution factor $D_{q\overline{q}}(\hat{s}=m_Z^2,Y)$
for $\cPqu\cPaqu$ (top, red boxes) and $\cPqd\cPaqd$ (bottom, blue crosses) production
as a function of the dilepton rapidity $Y$. The prediction from the \textsc{Pythia} simulation
(boxes and crosses) of the $q\overline{q}\to \gamma^*/\cPZ\to \mu^-\mu^+$ process and
the analytical distributions (solid curves) from Eq.~(\ref{eq:dilution-amplitude-factor})
are shown.
\label{fig:dilution}
}
\end{figure}

\begin{figure}[htbp]
\begin{center}
\includegraphics[width=\cmsfigwid]{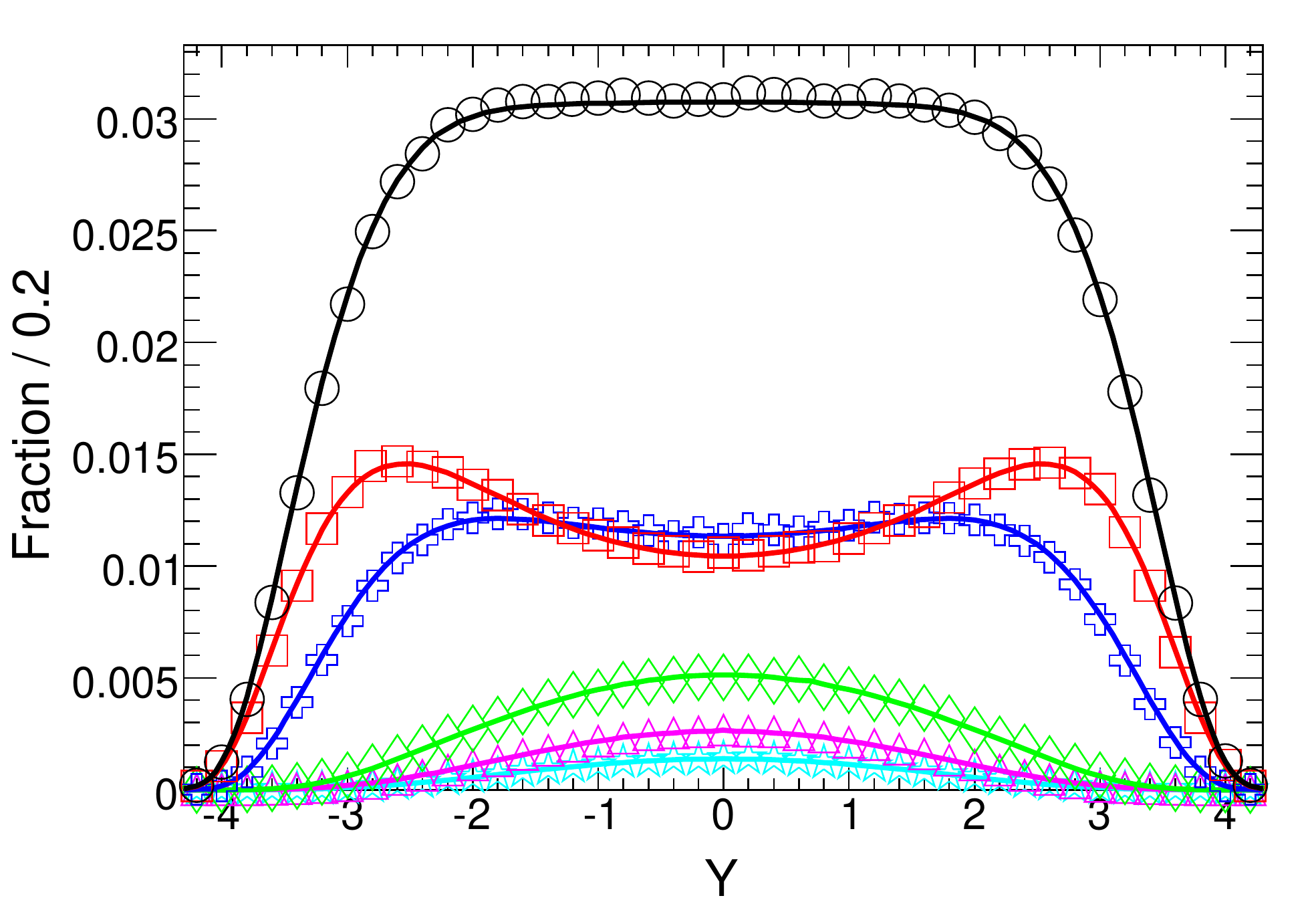} \\
\includegraphics[width=\cmsfigwid]{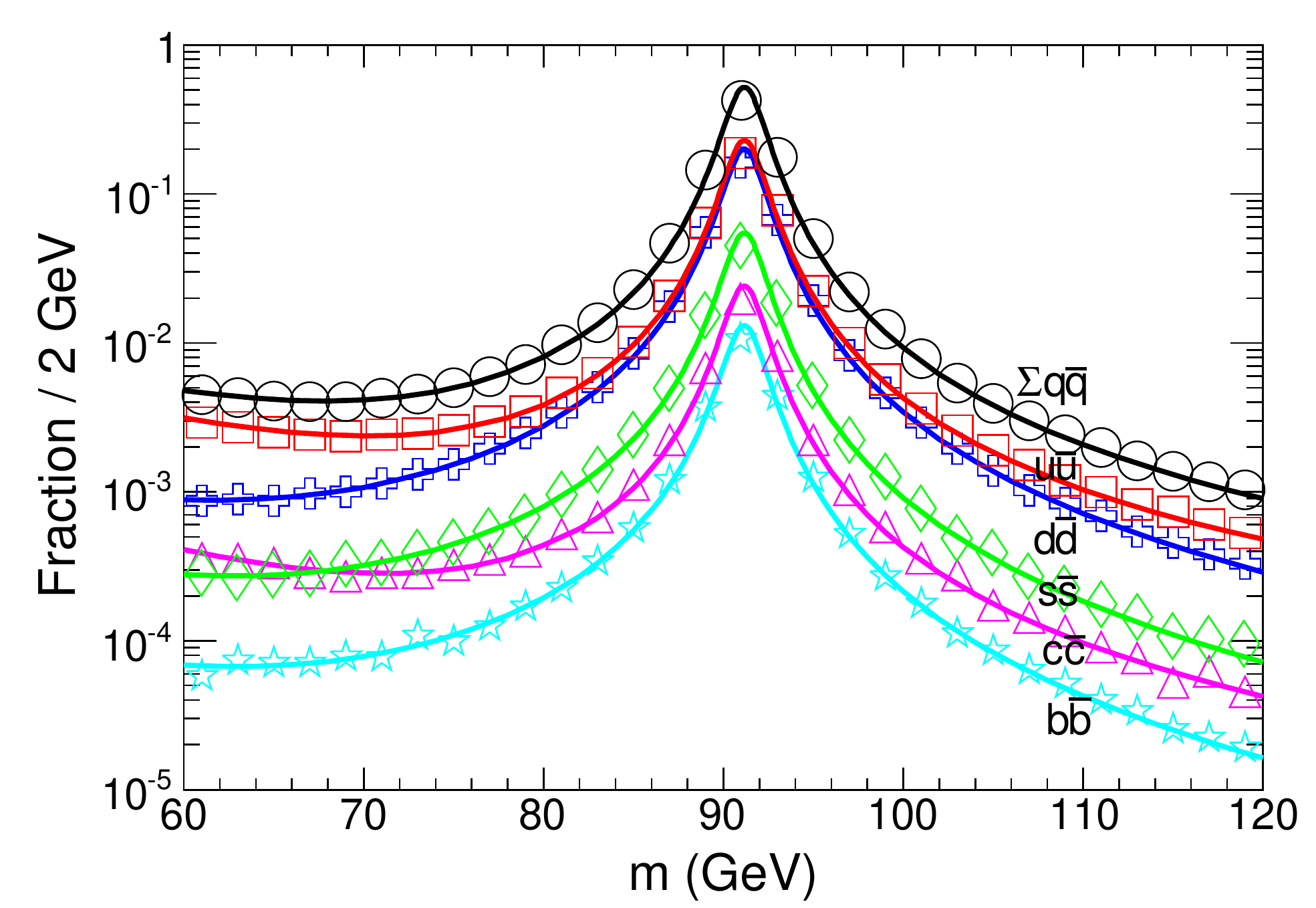} \\
\includegraphics[width=\cmsfigwid]{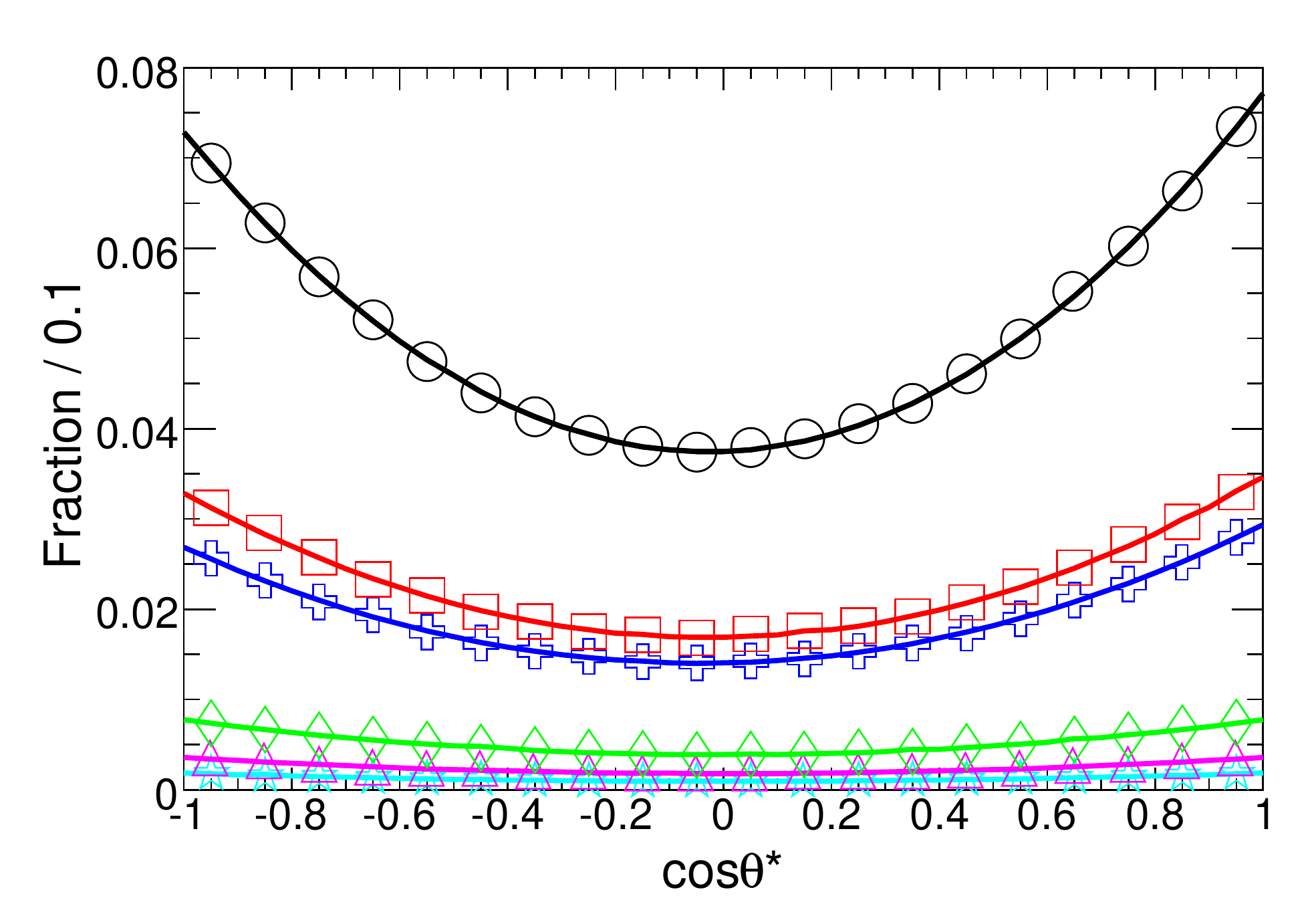}
\end{center}
\caption{
Distributions of $Y$ (top),  $m$ (middle), and $\cos\theta^\ast$ (bottom), from
\textsc{Pythia} simulation (points) of the $q\overline{q}\to \gamma^*/\cPZ\to \mu^-\mu^+$
process and analytical distributions from Eq.~(\ref{eq:dilution-amplitude-3}).
Distributions for five quark flavors are shown combined $\Sigma{q}\bar{q}$ (black circles)
and separately, in order of decreasing contribution:
$\cPqu\cPaqu$ (red boxes),
$\cPqd\cPaqd$ (blue crosses),
$\cPqs\cPaqs$ (green diamonds),
$\cPqc\cPaqc$ (magenta triangles),
and
$\cPqb\cPaqb$ (cyan stars).
Distributions are normalized to unit area and are shown as fractions of events per bin.
\label{fig:analytical}
}
\end{figure}

The factors $F_{q\overline{q}}(\hat{s},Y)$ and $D_{q\overline{q}}(\hat{s},Y)$ both arise from the PDFs
and can be extracted from parameterizations such as in
Refs.~\cite{CTEQ6:2004, Lai:2010, Martin:2009iq, Ball:2010de}.
We choose to extract PDFs numerically from the
leading-order (LO) parameterization \textsc{cteq6}~\cite{CTEQ6:2004} to match the LO
model of the process. We parameterize the PDFs analytically using polynomial and
exponential functions in the relevant range of $x$ and with
coefficients that are also functions of $\hat{s}$.
The relevant range of $x$ for this analysis is $1.1\times10^{-3}<x<1.4\times10^{-1}$,
motivated by the detector acceptance and the dilepton
mass selection criteria presented in Section~\ref{sec:method}.
This analytical parameterization is illustrated in Fig.~\ref{fig:pdfs}.
We choose an analytical parameterization because of the computational speed and
ease of implementation. Its performance has been
cross-checked with numerical computations. For systematic uncertainty studies,
we check the performance against simulations with other PDF
models~\cite{DeRoeck:2011na, Botje:2011sn}, such as next-to-leading-order (NLO)
\textsc{cteq}~\cite{Lai:2010}, \textsc{mstw}~\cite{Martin:2009iq}, and \textsc{nnpdf}~\cite{Ball:2010de}.

The function $F_{q\overline{q}}(\hat{s},Y)$ is the effective cross section factor that scales
the elementary parton-level cross section. This factor quickly decreases as the
energy scale approaches values comparable to the full proton energies,
as illustrated in Fig.~\ref{fig:factor} for $Y=0$ production.
The factor $D_{q\overline{q}}(\hat{s},Y)$ reflects the fact that the quark direction is generally
unknown and is taken as the boost direction of the dilepton system, because of the higher
probability for valence quarks to provide the boost.
For $q=u$ or $d$ this factor ranges between 0 and 1 as $|Y|$ changes from 0 to 4,
as illustrated in Fig.~\ref{fig:dilution} for $\hat{s}$ around the $\cPZ$ pole.
From Eq.~(\ref{eq:dilution-amplitude-factor}) it follows that $D_{q\overline{q}}=0$ for $q=\cPqs,\cPqc,\cPqb$
under the assumption $f_{\bar{q}}(x,Q^2) = f_{q}(x,Q^2)$, which is a good
approximation in the current PDF model.
A challenge at the LHC is that the dilution factor $D_{q\overline{q}}$ is small
for the typical range of $Y$ values in the detector acceptance region, as discussed below.
Information about $\sin^2\theta_{\PW}$ or individual fermion couplings
is contained in the shape of the three-dimensional distributions
in Eq.~(\ref{eq:dilution-amplitude-3}) and enters through the elementary couplings of the
electroweak bosons and fermions in the process $q\overline{q}\to\gamma^*/\cPZ\to \ell^-\ell^+$.

Figure~\ref{fig:analytical} illustrates the projections of the differential cross section from
Eq.~(\ref{eq:dilution-amplitude-3}) in $Y$,  $m=\sqrt{\hat{s}}$, and $\cos\theta^\ast$
for the five different quark flavors and combined.
The relative fractions of the different quark flavors are 0.450, 0.375, 0.103, 0.048, and 0.025
for $\cPqu\cPaqu$, $\cPqd\cPaqd$, $\cPqs\cPaqs$, $\cPqc\cPaqc$, and $\cPqb\cPaqb$, respectively.
The results of the analytical model leading to Eq.~(\ref{eq:dilution-amplitude-3}) show
good agreement with the predictions from conventional LO numerical
Monte Carlo (MC) simulation using the \textsc{Pythia} generator~\cite{Sjostrand:2006}
with LO \textsc{cteq6}~\cite{CTEQ6:2004} PDFs.
The above cross section is parameterized at leading order in both strong (QCD)
and electroweak (EWK) interactions. Effects from NLO QCD
contributions are studied with a detailed NLO \textsc{Powheg}~\cite{Nason:2004,Frixione:2007,Alioli:2008}
simulation, which includes contributions from both initial-state gluon radiation and quark-gluon scattering.
Effects from NLO EWK contributions are expected to be small compared to the precision
of this analysis. EWK corrections are absorbed in a definition of the effective
weak mixing parameter, $\sin^2\theta_\text{eff}$~\cite{Nakamura:2010zzi}.
We use $\sin^2\theta_\text{eff}$ in place of $\sin^2\theta_{\PW}$ for the rest of this paper.

We apply the above technique to the measurement of the weak mixing angle.
We take the SM description of electroweak interactions and PDFs
in the proton as given, and allow only the effective weak mixing
angle $\theta_\text{eff}$ to be unconstrained. More generally, the above formalism
with the multivariate analysis of the Drell--Yan process allows us to study
the elementary couplings of fermions to electroweak neutral fields, such as
$\gamma^{*}/\cPZ$ in the SM, as well as the structure functions of the proton.

\section{Detection of the Drell--Yan Events with CMS}
\label{sec:reco}

We apply the above method in an analysis of the $q\overline{q}\rightarrow \gamma^{*}/\cPZ \rightarrow\mu^-\mu^+$
process and measure $\sin^2\theta_\text{eff}$. The choice of $\mu^-\mu^+$, as opposed to $e^-e^+$,
is motivated by the more reliable description of the detector and background effects,
as well as the fact that this final state has not yet been studied for $\sin^2\theta_\text{eff}$
measurements in $q\overline{q}$ interactions. However, we do not expect any limitations
in the method for future application to other final states.
The expected multivariate distributions in Eq.~(\ref{eq:dilution-amplitude-3}) are
modified by smearing due to detector resolution and photon final-state radiation (FSR),
and by acceptance effects and non-uniform reconstruction as a function of the observables.
All these effects are taken into account in the analytical parameterization, as shown below.

A detailed description of the CMS detector can be found in Ref.~\cite{CMS:2010}.
The central feature of CMS is a 3.8 T superconducting solenoid of 6 m internal diameter.
Within the field volume are the silicon pixel and strip tracker,
the crystal electromagnetic calorimeter (ECAL), and the brass/scintillator hadron calorimeter (HCAL).
This analysis of the dimuon final state does not rely strongly on ECAL or HCAL measurements.
Muons are measured in the window $|\eta|<2.5$ with the tracker and muon system.
The pseudorapidity $\eta$ is defined as $\ln\cot(\theta/2)$ with the polar
angle $\theta$ measured in the laboratory frame.

The silicon tracking detector (tracker)~\cite{TDR}
consists of 1440 silicon pixel and 15\,148 silicon strip detector modules.
The pixel modules provide two-dimensional measurements of the hit position
in the module planes, which translate into three-dimensional
measurements in space, and are arranged in three layers in the barrel and
two layers in the forward regions. The silicon strip detector is composed
of 10 layers in the barrel region, four of which are double-sided,
and 12 layers in the endcap, where three out of six rings are with double-sided modules.
Precise determination of the position of all silicon modules (alignment) is one of
the critical aspects for achieving the designed resolutions of muon track parameters  and
is an important element of this analysis~\cite{Chatrchyan:2009sr}.
The muon system has detection planes composed of three distinct detector technologies
installed outside the solenoid and embedded in the steel return yoke:
drift tubes (in the barrel, $|\eta|<1.2$),
cathode strip chambers (in the endcaps, $0.9<|\eta|<2.5$),
and resistive plate chambers
(in both barrel and endcap regions, $|\eta|<1.6$)~\cite{TDR}.

This analysis uses data from proton-proton collisions at $\sqrt{s}$ = 7 TeV collected
during 2010 and 2011, and corresponding to ($1.07\pm0.05\fbinv$)  of integrated luminosity.
The signal and background processes $q\overline{q}\to\gamma^*/\cPZ\rightarrow\mu^-\mu^+$ and
$\tau^-\tau^+$ have been simulated with the NLO QCD generator \textsc{Powheg}.
Parton showering is simulated using \textsc{Pythia}. The NLO PDFs used are \textsc{ct10}~\cite{Lai:2010}.
Background samples of $\PW$+jets and $\cPqt\cPaqt$ are generated using \textsc{MadGraph}~\cite{Alwall:2007},
\textsc{Pythia}, and \textsc{Tauola} \cite{Davidson:2010}.
Backgrounds from $\PW\PW$, $\PW\cPZ$, $\cPZ\cPZ$, and QCD are generated using \textsc{Pythia}.
Generated events are processed through the CMS detector simulation and reconstruction.
The detector simulation is based on \textsc{Geant4}~\cite{Sulkimo:2003, Allison:2006}.

Muon candidates are selected from a sample triggered online by events with at least two
muons within the volume defined by $|\eta|<2.4$ and with transverse momentum
($p_{\sss T}$) requirements.  These requirements depend on the period of data-taking; however,
they always accept two muons with $p_{\sss T}$ of at least 8 and 13\GeV, respectively.
Offline, muon tracks are first reconstructed independently in the tracker and the muon system.
Muon candidates are then reconstructed by two different algorithms~\cite{CMS-PAS-MUO-10-002}.
The global muon algorithm matches tracks in the tracker to tracks in the muon system,
and then refits the individual hits in the tracker and muon system to one overall track.
The tracker muon algorithm extrapolates tracks in the tracker with $p_{\sss T}>0.5\GeV$
and $p>2.5\GeV$ to the muon system, and a track is taken to be a muon candidate if it matches
at least one track segment in the muon system. Both algorithms take into account energy
loss and multiple scattering in the steel yoke of the CMS magnet.
Selection criteria demand at least 10 hits in the tracker, including one in the pixel detector,
at least one hit in the muon system, and a normalized $\chi^2<10$ for the global fit.

Muons are required to have a small impact parameter, less than 2 mm measured with
respect to the beam spot in the plane perpendicular to the beam direction.
This requirement removes cosmic-ray muons and background
events with displaced vertexes. We further require the angle between the two muon
tracks to be larger than 2.5 mrad in the laboratory frame when the direction
of one of the tracks is reversed.
This removes any remaining cosmic-ray background and has negligible
effect on the signal. To isolate single muons from muons overlapping with jets,
the sum of the transverse momentum of tracks in the tracker (excluding the muon in question)
within a surrounding cone of $\Delta R \equiv \sqrt{(\Delta\eta)^2+(\Delta\phi)^2} <0.3$
is required to be less than 15\% of the measured transverse momentum of the muon,
where $\Delta\eta$ and $\Delta\phi$ are the differences in pseudorapidity and
in azimuthal angle in radians between the muon and the track.
The ECAL and HCAL are not used for muon isolation, to reduce the effect from
FSR and to maximize the amount of signal events.

The kinematic requirements in the laboratory frame are $|\eta|<2.4$ and $p_{\sss T}>18$
and $8$ GeV for the two muons. We introduce additional requirements
in the Collins--Soper frame in order to simplify the acceptance parameterization:
$|\eta^*|<2.3$ and $p^*_{\sss T}>18$ GeV, where $\eta^*$ and $p^*_{\sss T}$
are defined with respect to the $z^\prime$-axis, described previously.
We also require the dimuon transverse momentum in the laboratory frame to be less
than 25 GeV in order to suppress the contribution of events with hard jet radiation.
Dilepton events are selected from events containing two oppositely charged, isolated,
high-$p_{\sss T}$ muons with a dilepton invariant mass $m$ in the range $80-100$ GeV.
The dimuon rapidity $Y$ is calculated from the lepton four-momenta as shown in Eq.~(\ref{eq:rapidity}).
Restrictions on $\theta^{*}$ and $Y$ are motivated by detector acceptance effects, as discussed
in Section~\ref{sec:method}. The number of selected events in the data is $N=297\,364$.

\section{Analysis Method}
\label{sec:method}

We use an unbinned extended maximum-likelihood fit that simultaneously describes
the signal and background yields and the parameters of the $(Y,\hat{s},\cos\theta^*)$
distributions. The likelihood function is written  as
\begin{widetext}
\begin{eqnarray}
{\mathcal{L}} =
\exp\left( - n_\text{sig}-n_\text{bkg}  \right)
\prod_i^{N} \left(
n_\text{sig} \times\mathcal{P}_\text{sig}(\vec{x}_{i}; \theta_\text{eff}; \vec{\xi})
+n_\text{bkg} \times\mathcal{P}_\text{bkg}(\vec{x}_{i};~\vec{\xi})
\right)\,,
\label{eq:likelihood}
\end{eqnarray}
\end{widetext}
where each event candidate $i$ is characterized by a set of three observables
$\vec{x}_{i}=\{Y,\hat{s},\cos\theta^*\}_i$,
$n_\text{sig}$ is the number of signal events, which includes
all intermediate states ($\gamma^*$, $\cPZ$, and their interference),
$n_\text{bkg}$ is the small number of background events,
$\mathcal{P}_\text{sig}(\vec{x}_{i};\vec{\xi})$ and $\mathcal{P}_\text{bkg}(\vec{x}_{i};\vec{\xi})$
are the probability density functions for signal and background processes,
and $\vec{\xi}$ represent the parameters of these functions.
The signal probability density function is defined as
\begin{widetext}
\begin{eqnarray}
\mathcal{P}_\text{sig}(Y, \hat{s}, \cos\theta^\ast; \theta_\text{eff}) =
\mathcal{G}(Y, \hat{s}, \cos\theta^\ast) \times
\int_{-\infty}^{+\infty} dx \,\mathcal{R}(x) \,
\mathcal{P}_\text{ideal}(Y, \hat{s}-x, \cos\theta^\ast; \theta_\text{eff})
\,.
\label{eq:signal-pdf}
\end{eqnarray}
\end{widetext}
The ideal distribution $\mathcal{P}_\text{ideal}(Y, \hat{s}, \cos\theta^\ast; \theta_\text{eff})$
in Eq.~(\ref{eq:signal-pdf}) is the Drell--Yan cross section defined in Eq.~(\ref{eq:dilution-amplitude-3}).
We correct $\mathcal{P}_\text{ideal}$ for detector effects, such as acceptance,
parameterized with $\mathcal{G}(Y, \hat{s}, \cos\theta^\ast)$,
and resolution and photon emission (FSR), parameterized with $\mathcal{R}(x)$
where $x$ is the change in the dilepton center-of-mass energy squared.

\begin{figure}[htbp]
\begin{center}
\includegraphics[width=\cmsfigwid]{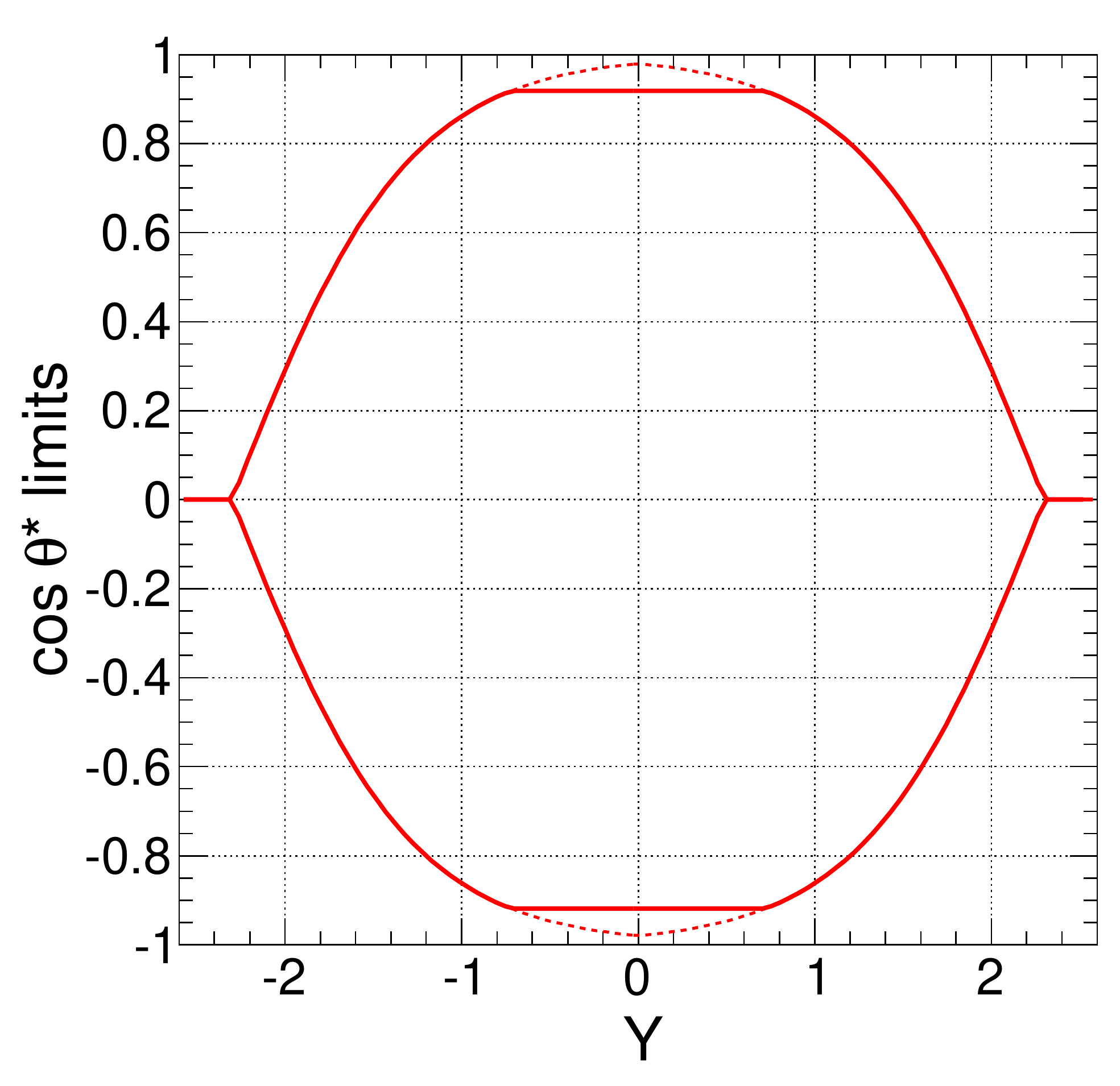}
\end{center}
\caption{
Accepted $\cos\theta^\ast$ range as a function of $Y$ for $\hat{s}=m_Z^2$ and
for the kinematic selection used in this analysis. The outer boundary corresponds
to Eq.~(\ref{eq:acceptance1}) and the horizontal lines near $\cos\theta^*=\pm0.92$
correspond to Eq.~(\ref{eq:acceptance2}).
\label{fig:acceptance}
}
\end{figure}

The acceptance function $\mathcal{G}(Y, \hat{s}, \cos\theta^\ast)$ describes the
nonuniform reconstruction efficiency as a function of the three observables,
which includes effects from online trigger requirements, detector acceptance,
reconstruction algorithms, and selection requirements.
The most important effect is the loss of particles near the beam directions
and the second-most-important effect is the minimum transverse momentum
requirement on the leptons. Otherwise, the efficiency across the acceptance range,
defined by the selection requirements $|\eta^*|<Y_\text{max}=2.3$
and $p^*_{\sss T}>p_\text{min}=18\GeV$, is close to uniform.
The above selection requirements define a sharp boundary in
$(Y,\hat{s},\cos\theta^*)$ space, which can be expressed as limits on
$\cos{\theta^*}$ for given $Y$ and $\hat{s}$ values as follows:
\begin{eqnarray}
&& |\cos{\theta^*}|<\tanh(Y_\text{max}-|Y|) \,,
\label{eq:acceptance1} \\
&& |\cos{\theta^*}|<\sqrt{1-4p_\text{min}^2/\hat{s}} ~.
\label{eq:acceptance2}
\end{eqnarray}
This boundary is illustrated in Fig.~\ref{fig:acceptance} in the $(Y, \cos\theta^\ast)$ plane
for a fixed value $\hat{s}=m_Z^2$.

\begin{figure}[htbp]
\begin{center}
\includegraphics[width=\cmsfigwid]{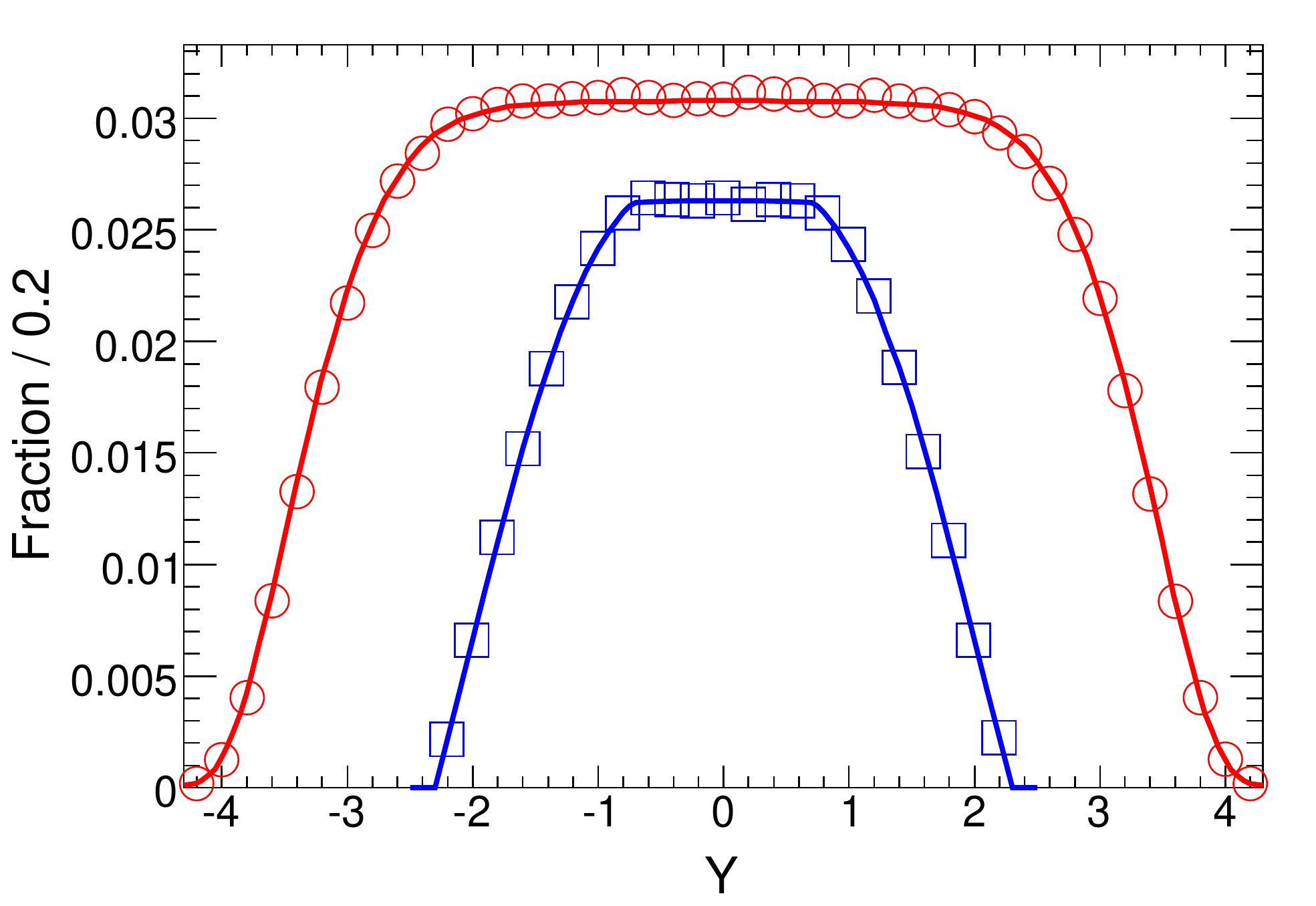} \\
\includegraphics[width=\cmsfigwid]{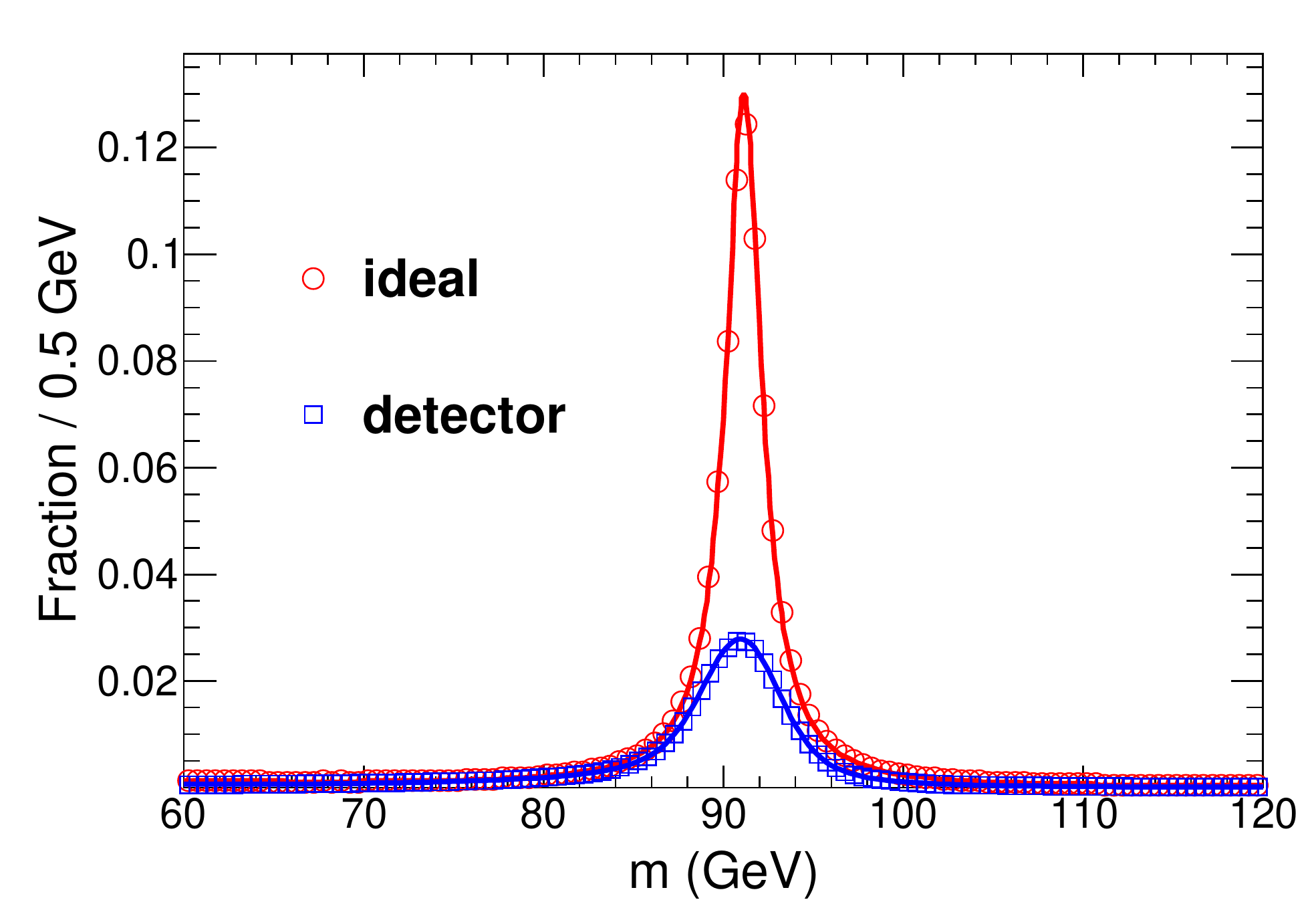} \\
\includegraphics[width=\cmsfigwid]{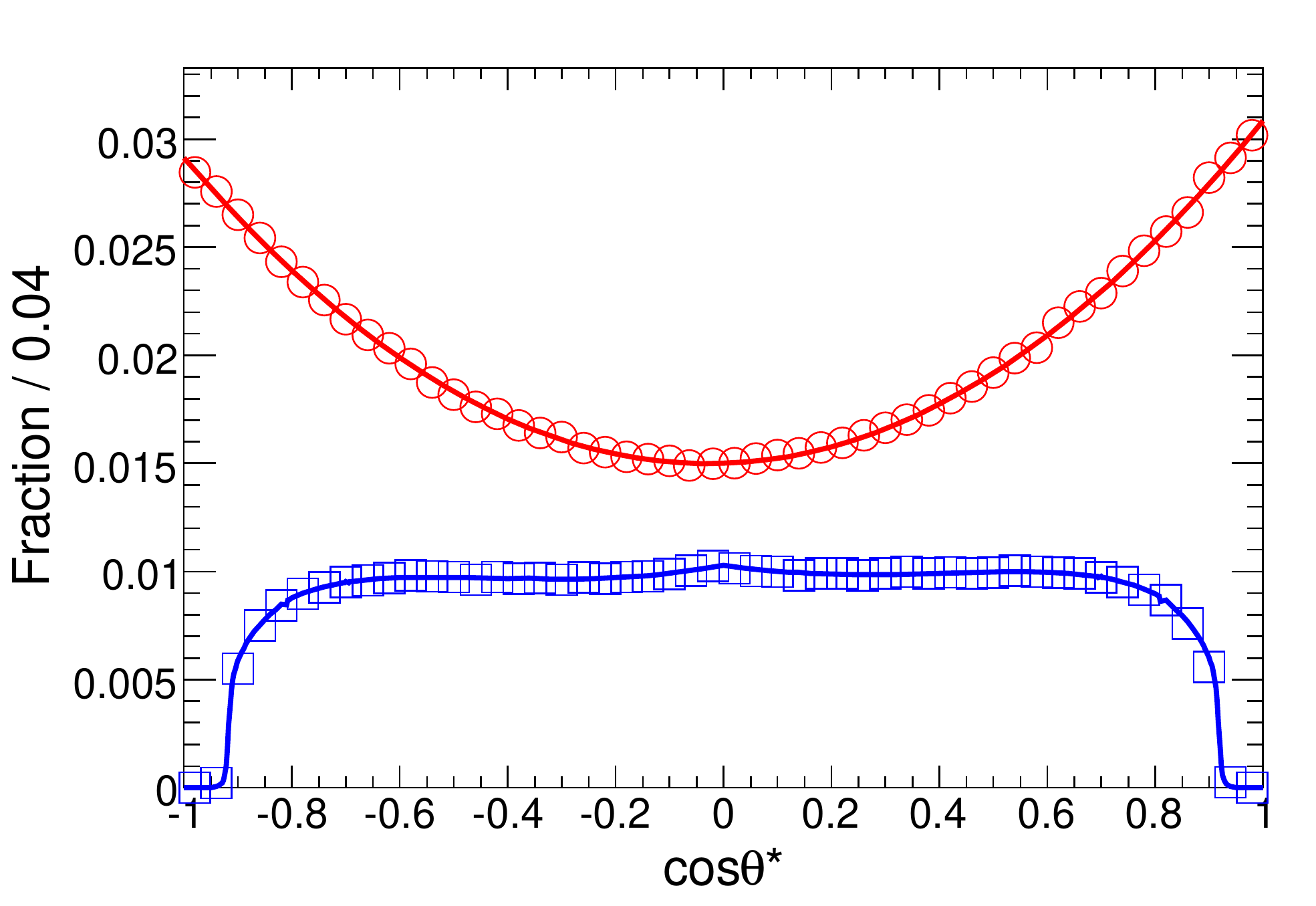}
\end{center}
\caption{
Distributions of $Y$ (top),  $m$ (middle), and $\cos\theta^\ast$ (bottom), from \textsc{Pythia} simulation
(points) of the $q\overline{q}\to \gamma^*/\cPZ\to \mu^-\mu^+$ process and its analytical
parameterization (smooth curve).
Combined distributions from Fig.~\ref{fig:analytical} appear in red at the top of each plot
(red circles for an ``ideal" simulation),
while distributions after acceptance and resolution effects, including photon FSR,
appear in blue below (blue squares for a simplified ``detector" simulation).
Distributions are normalized to unit area and are shown as fractions of events per bin.
\label{fig:detector-effects}
}
\end{figure}

The effect of smearing the muon track parameters, such as the muon momentum and
direction, due to detector resolution and FSR, is most evident in the invariant mass distribution.
This effect is parameterized with the function $\mathcal{R}(x)$ in Eq.~(\ref{eq:signal-pdf}).
Both acceptance and resolution effects are illustrated in Fig.~\ref{fig:detector-effects}, where
the analytical parameterization of $Y$,  $m=\sqrt{\hat{s}}$, and $\cos\theta^\ast$ is in good agreement
with LO simulation in both QCD and EWK, as generated by \textsc{Pythia}.
Although a wider $m$ range is investigated, the analysis is performed in the range
$80<m<100$~GeV to reduce uncertainties from FSR.
In this illustration, FSR is included and the major detector effects are introduced in the following way:
for the three track parameters ($p_{\sss T}$, $\phi$, $\theta$), we apply Gaussian
random smearing with standard deviation of
$\Delta p_{\sss T}=0.025 p_{\sss T}+0.0001 p_{\sss T}^2$ (with $p_{\sss T}$ in GeV),
$\Delta\phi=\Delta\theta=0.001$ rad, and neglect resolution effects on the track origin.
This simplified simulation of detector effects is found useful to isolate production model
uncertainties from the detector effect parameterization.

Further studies are performed with full \textsc{Geant4}-based modeling of the CMS detector
using the \textsc{Powheg} simulation of the dimuon events and with \textsc{Pythia} simulation
of the parton showering and FSR. In the parameterization of the acceptance function
$\mathcal{G}(Y, \hat{s}, \cos\theta^\ast)$, we model the small deviations from a uniform efficiency
with empirical polynomial functions that include correlations of the two observables
within the boundaries of the $(Y, \cos\theta^\ast)$ plane defined above.
This efficiency parameterization is derived from the simulation
with a fit where the parameters of the polynomial functions are left unconstrained.
The main effect is a loss of efficiency in the vicinity of the acceptance boundaries.
A similar approach is later employed as part of the systematic uncertainty studies
where the parameters of the efficiency model are left free in the fit to data.

In the parameterization of the resolution function $\mathcal{R}(x)$, FSR is modeled
with \textsc{Pythia} and resolution effects are taken from the full CMS detector simulation,
including the effects of tracker alignment on the tracking resolution. The function
$\mathcal{R}(x)$ is approximated with a sum of four Gaussian functions, to allow for the
analytical convolution in Eq.~(\ref{eq:signal-pdf}) and be flexible enough to
describe both detector resolution and FSR effects. Parameters of the $\mathcal{R}(x)$
function are left free in the fit to the simulated MC sample. The overall shift of the $\cPZ$ mass
in the resolution function $\mathcal{R}(x)$ is left free in the fit to data, effectively allowing
the energy scale to be determined from the data.

The background contribution is estimated by MC simulation;
the QCD component has been cross-checked with data.
The total expected background is about 0.05\% of the signal yield.
The background consists of the crossfeed from the $q\overline{q}\to \cPZ/\gamma^*\to \tau^+\tau^-$
process, QCD, $\cPqt\cPaqt$, and diboson production in nearly equal contributions.
The probability density function for the background $\mathcal{P}_\text{bkg}(\vec{x}_{i};~\vec{\xi})$
is parameterized in a similar manner to that shown in Eq.~(\ref{eq:signal-pdf}) with
an acceptance range defined by Eqs.~(\ref{eq:acceptance1}) and~(\ref{eq:acceptance2})
and the distributions within the acceptance boundaries parameterized with an empirical polynomial function.
The number of background events $n_\text{bkg}$ is fixed to the expected value of 157 events.

\begin{figure}[htbp]
\begin{center}
\includegraphics[width=\cmsfigwid]{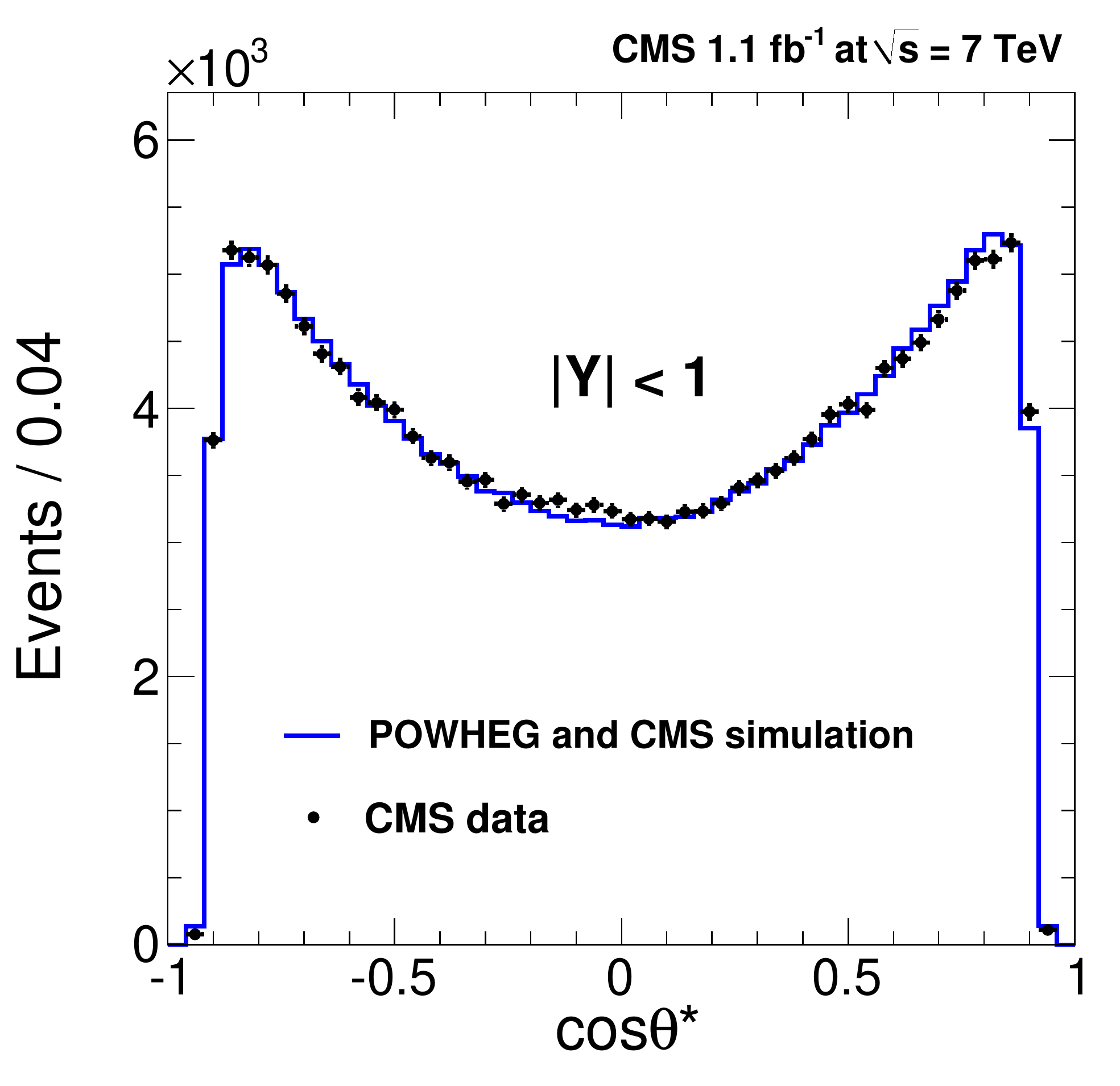}
\includegraphics[width=\cmsfigwid]{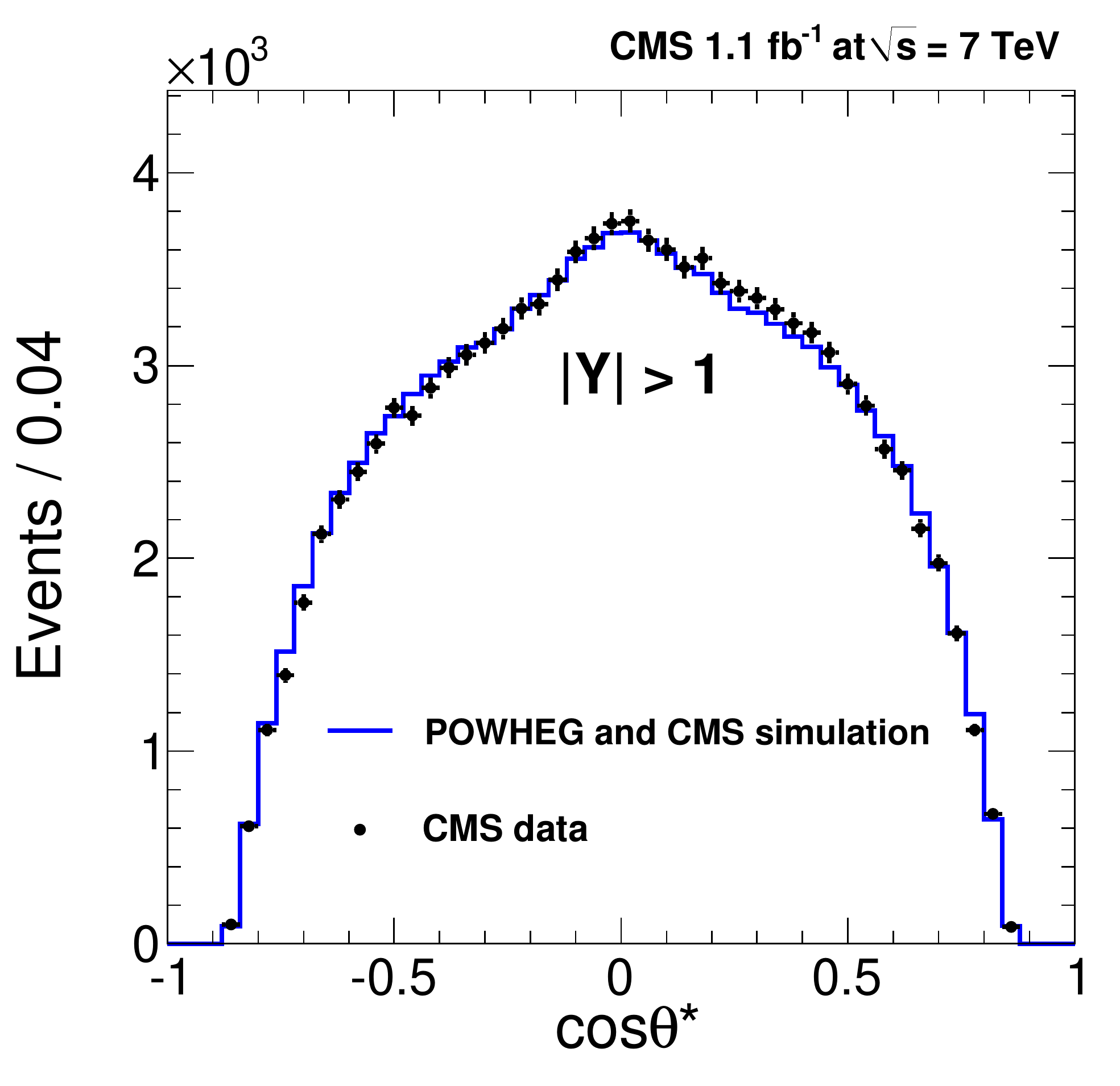}
\end{center}
\caption{
Distribution of $\cos\theta^\ast$ for data (points) and \textsc{Powheg}-based detector simulation
with $\sin^2\theta_\PW=0.2311$ (histogram) of the $q\overline{q}\to\gamma^*/\cPZ\to\mu^-\mu^+$
process for $|Y|<1$ (top) and  $|Y|>1$ (bottom).
\label{fig:HsY_data}
}
\end{figure}

In Fig.~\ref{fig:HsY_data} we show the $\cos\theta^\ast$ distributions in the data
separately for the $|Y|<1$ and  $|Y|>1$ regions, and compare them to the \textsc{Powheg}-based
simulation of the $q\overline{q}\to\gamma^*/\cPZ\to\mu^-\mu^+$ process in the CMS detector.
Together with Fig.~\ref{fig:dilution}, these distributions illustrate the challenge
of analyzing Drell--Yan events at the LHC. While the acceptance effects on the $\cos\theta^\ast$
distribution are moderate for smaller values of $|Y|$, the dilution is strong, as shown
in Fig.~\ref{fig:dilution}. In contrast, the larger values of $|Y|$ have a smaller
dilution effect, but the $\cos\theta^\ast$ range is strongly truncated because of
the limited acceptance, as shown in Fig.~\ref{fig:acceptance}. Therefore, the optimal
analysis of the angular distributions requires proper accounting for such correlations
among the three observables. At the same time, Fig.~\ref{fig:HsY_data} shows good
agreement between the data and MC simulation. Residual differences in the distributions
can be explained by the somewhat different value of $\sin^2\theta_\text{eff}=0.2311$
used in the simulation compared to the best value describing the data, and
by several systematic effects accounted for below, such as the tracker misalignment,
the momentum scale in the track reconstruction, and FSR modeling.

\begin{figure}[htbp]
\begin{center}
\includegraphics[width=\cmsfigwid]{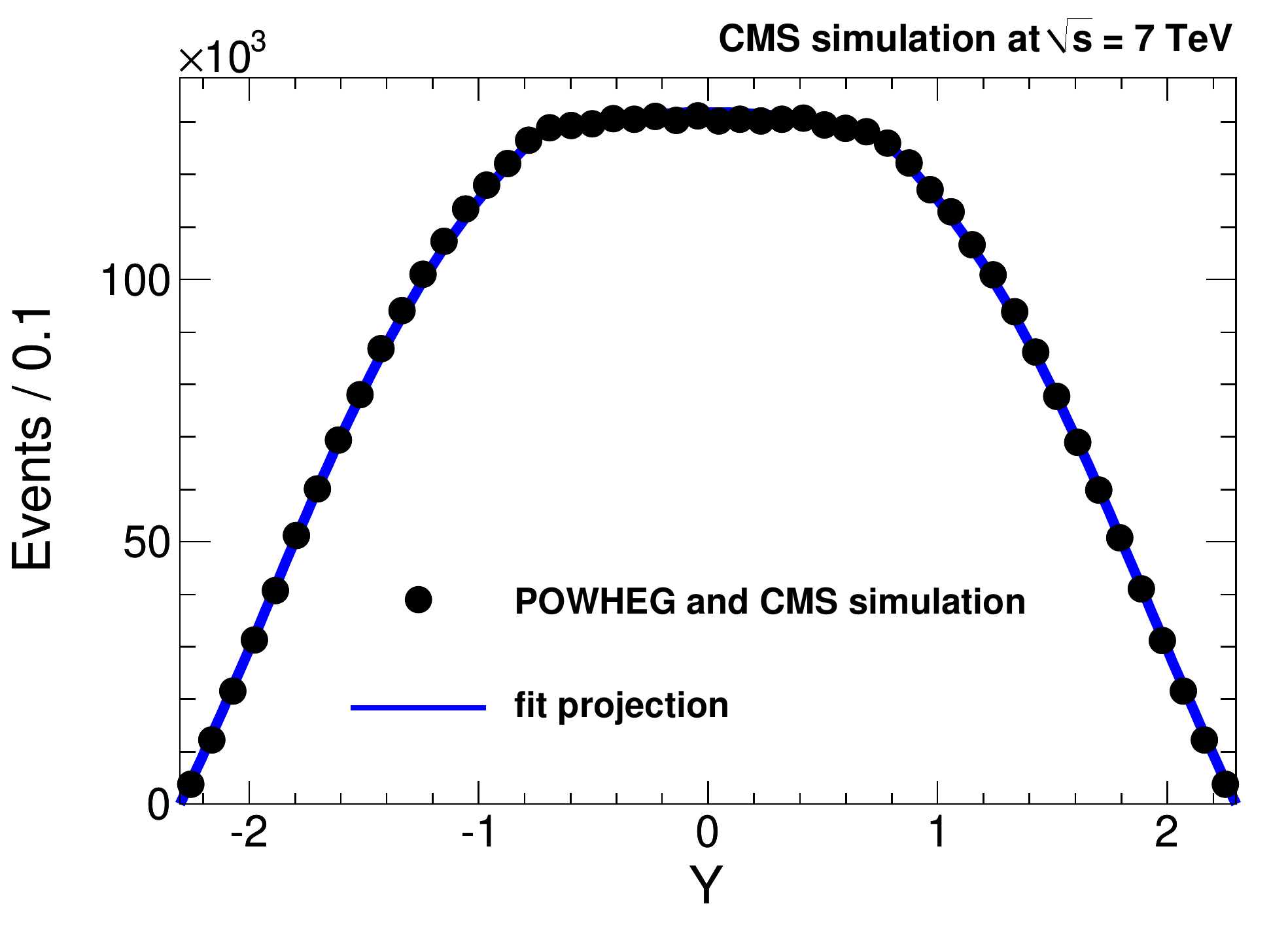} \\
\includegraphics[width=\cmsfigwid]{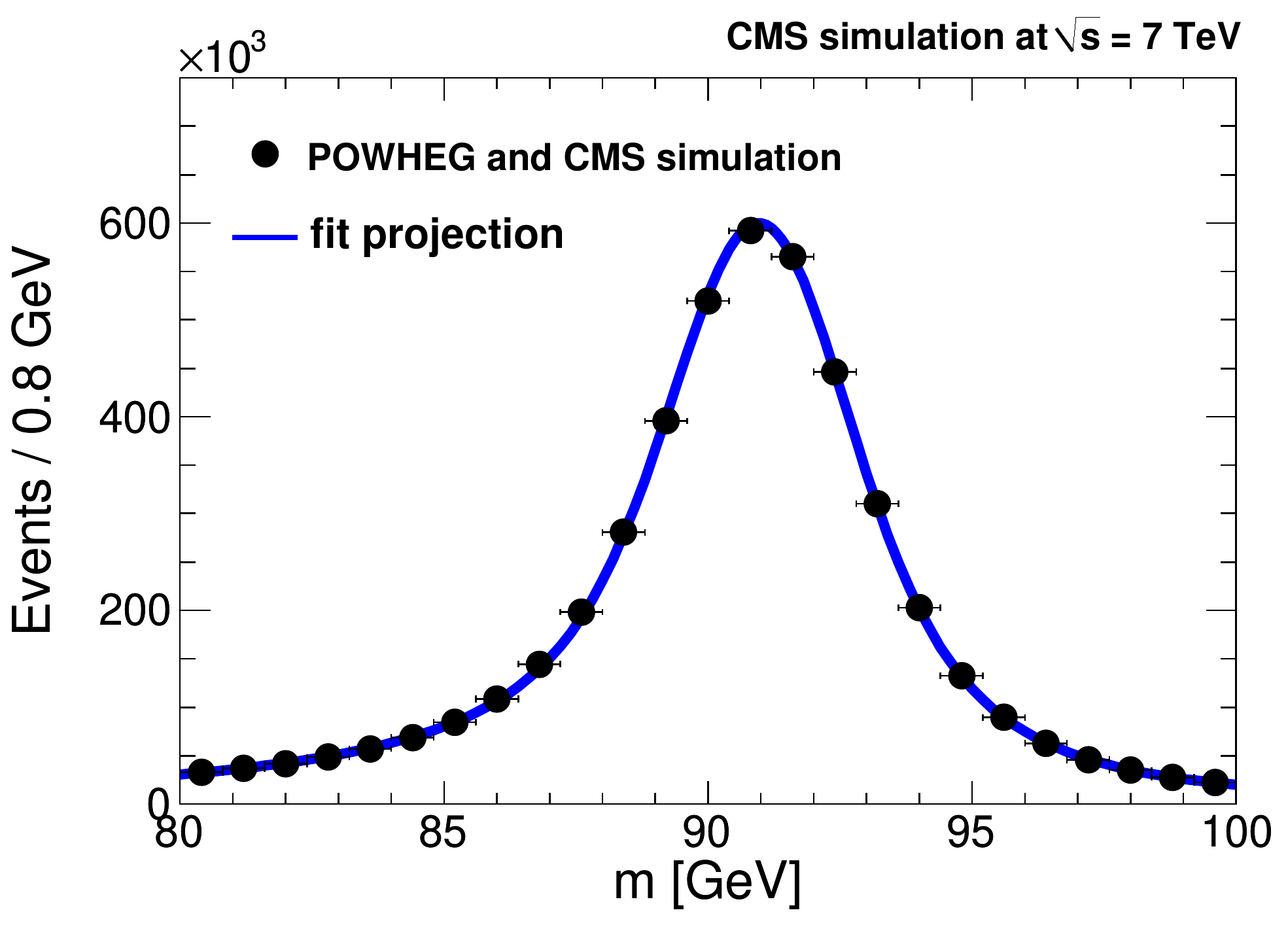} \\
\includegraphics[width=\cmsfigwid]{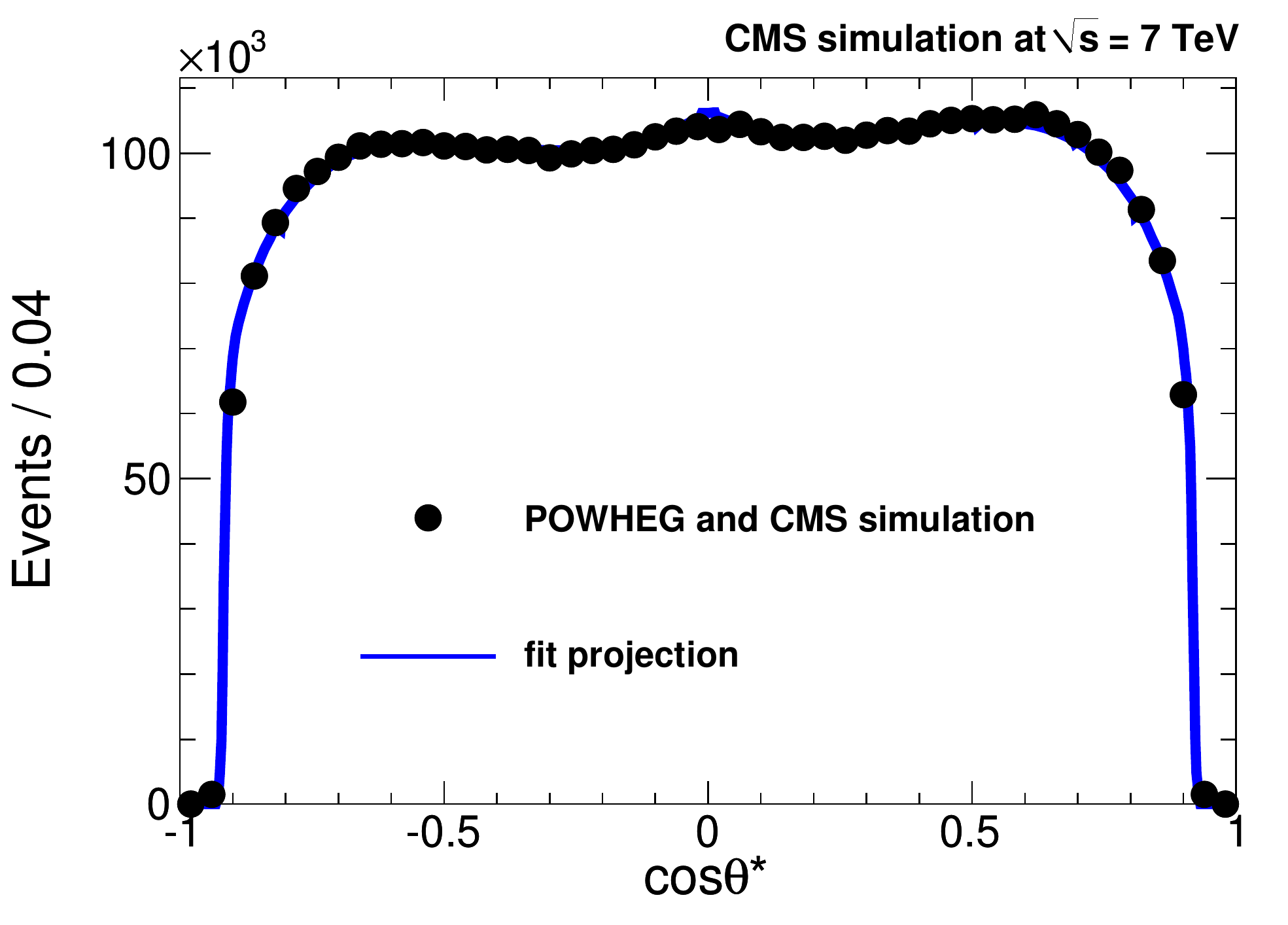}
\end{center}
\caption{
Distributions of $Y$ (top),  $m$ (middle), and $\cos\theta^\ast$ (bottom) from the \textsc{Powheg}-based
detector simulation with $\sin^2\theta_\PW=0.2311$ of the $q\overline{q}\to\gamma^*/\cPZ\to\mu^-\mu^+$
process (points). The MC sample corresponds to an integrated luminosity of 16\fbinv.
The lines show the projections of the probability density functions.
\label{fig:CmsMC}
}
\end{figure}

A ``blind" analysis of the data is performed, in which the fit result is not examined
until a review of the entire analysis is complete, including the evaluation of all associated
systematic uncertainties.
However, while the analysis is performed ``blind,"  the quality of the fits to the MC simulation
and data is examined. We test the performance of the fitting procedure using samples
generated using Monte Carlo simulations, with each separate sample containing the
same number of events observed in the data.
Signal events are generated with the \textsc{Powheg}-based CMS detector simulation
with an input value of $\sin^2\theta_\text{eff}=0.2311$.
The number of background events is Poisson distributed according to expectation.
After the corrections discussed below are applied, the pull distribution
is in agreement with a unit-width Gaussian distribution centered at zero.
A comparison of the MC sample projections and the probability
density functions are shown in Fig.~\ref{fig:CmsMC}.

\begin{figure}[htbp]
\begin{center}
\includegraphics[width=\cmsfigwid]{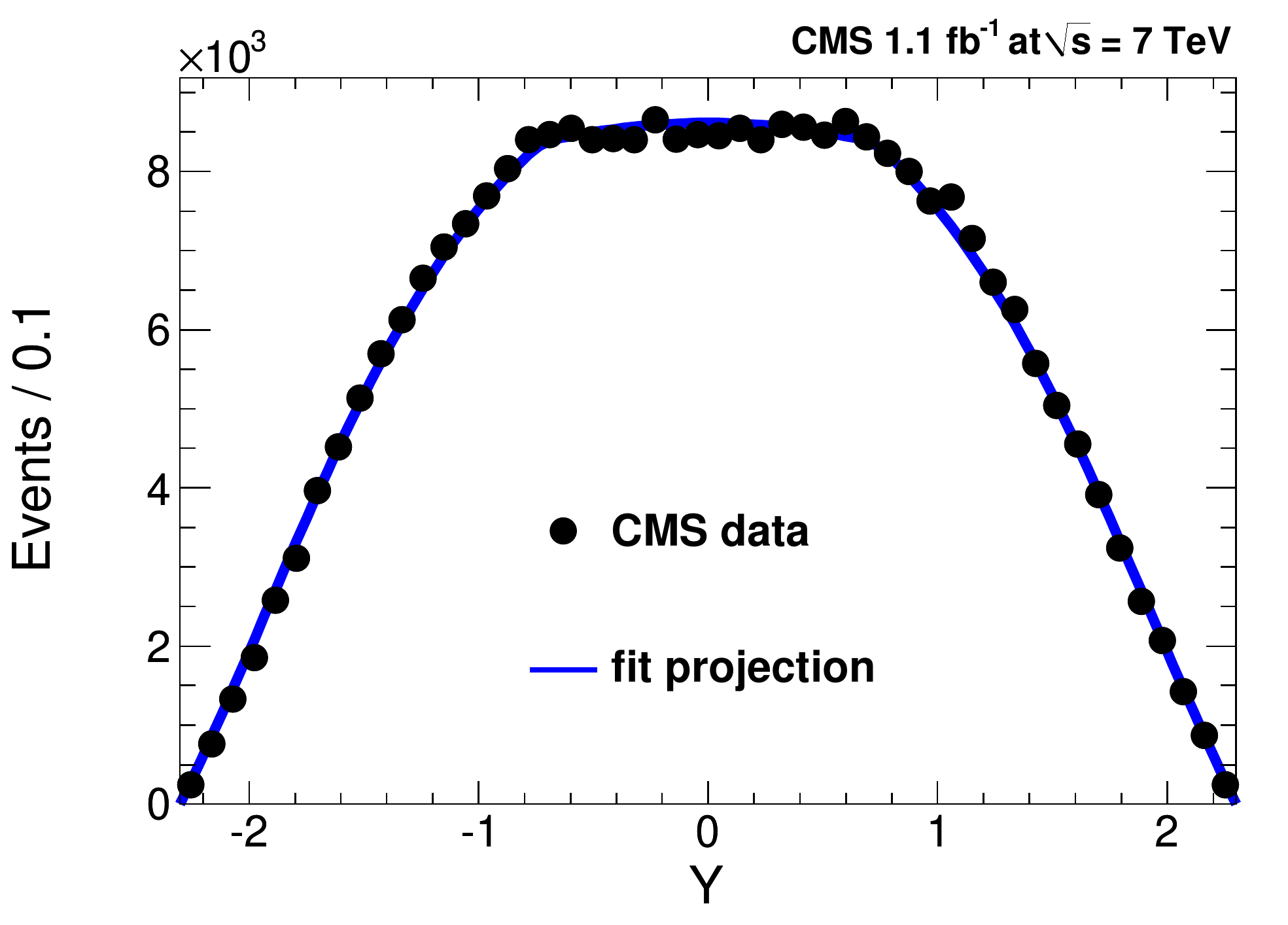} \\
\includegraphics[width=\cmsfigwid]{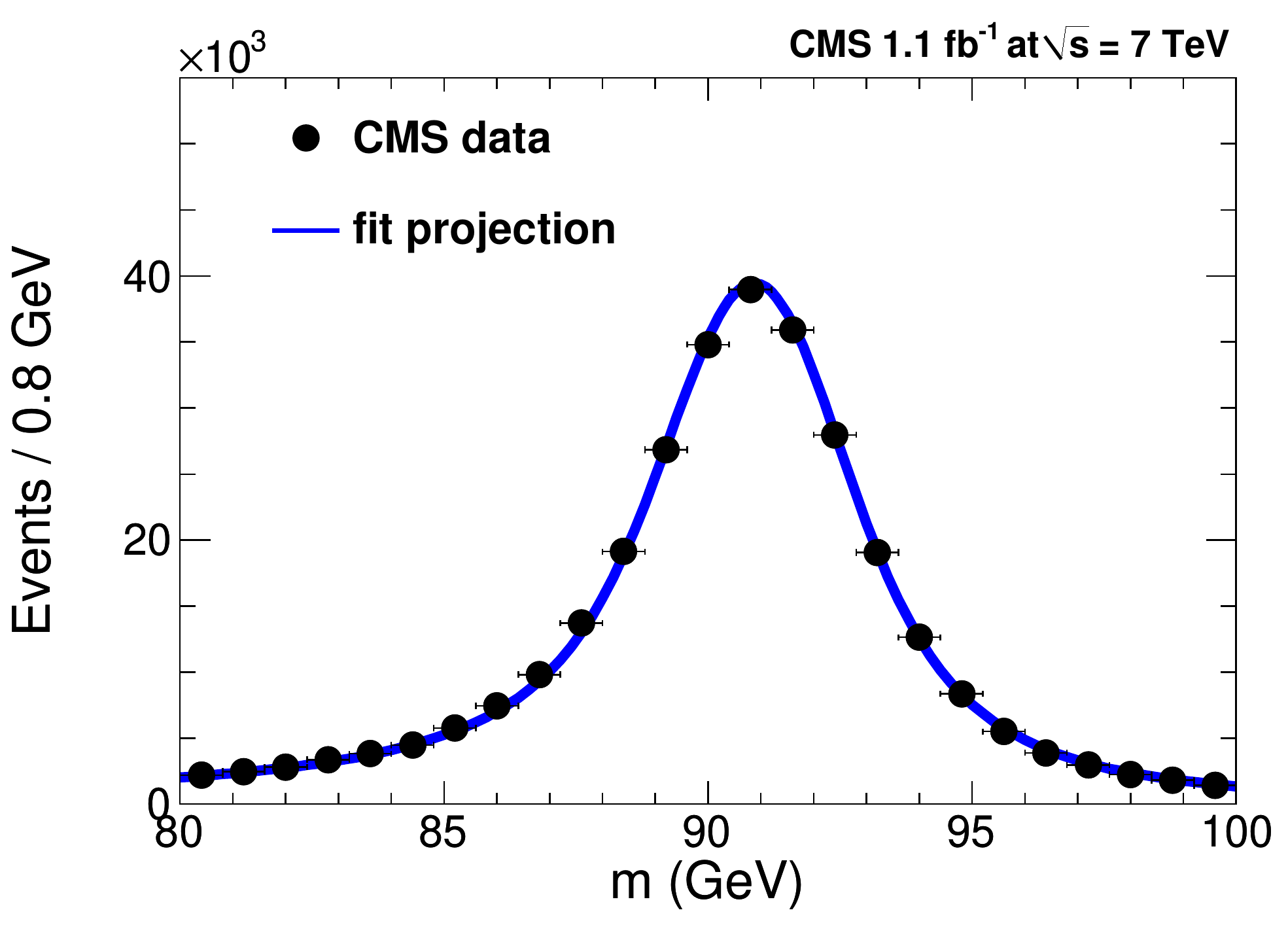} \\
\includegraphics[width=\cmsfigwid]{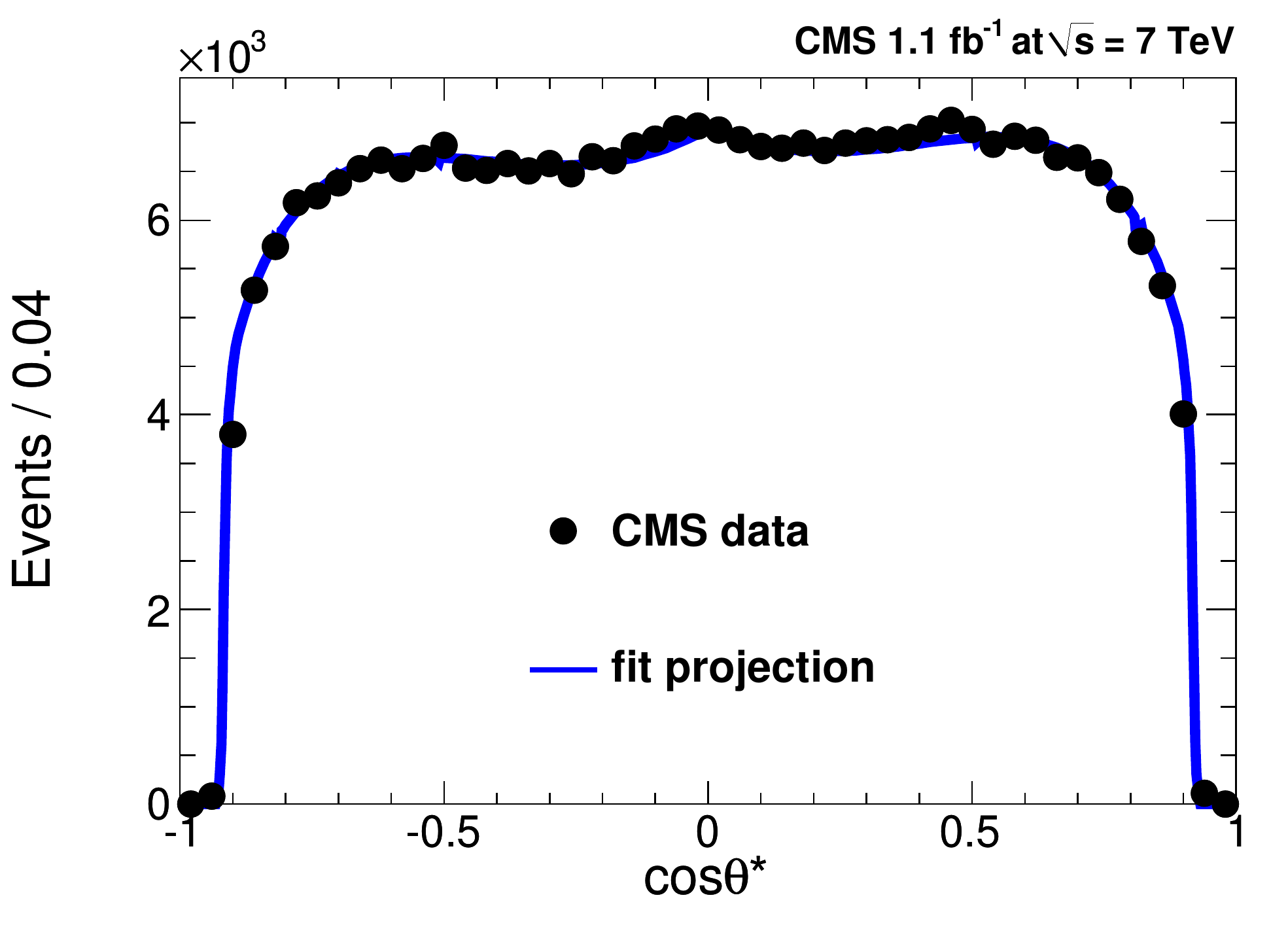}
\end{center}
\caption{
Distributions of $Y$ (top),  $m$ (middle), and $\cos\theta^\ast$ (bottom)
in the analysis of $q\overline{q}\to\gamma^*/\cPZ\to\mu^-\mu^+$ events from CMS (points).
The lines show the projections of the probability density functions.
\label{fig:CmsData}
}
\end{figure}

We examine the quality of the fit to the data by comparing the data distributions
to the likelihood model expectations, and comparing the fit likelihood value $\mathcal{L}$
and the observed statistical uncertainty to those expected with the generated samples.
Projections of the data and the probability density functions are shown in Fig.~\ref{fig:CmsData}.
They exhibit similar agreement as with the simulation shown in Fig.~\ref{fig:CmsMC}.
Correction for the energy scale is already included in the fit model.
The observed statistical error on $\sin^2\theta_\text{eff}$ of $\pm0.0020$ is in good
agreement with what is expected from the MC samples discussed above.
We find only small differences when comparing the likelihood value $\mathcal{L}$ for
generated experiments from the likelihood model, the \textsc{Powheg}-based
CMS detector simulation, and the data. The level of agreement is consistent with
typical differences due to imperfect efficiency function modeling and NLO effects
discussed below. The variations do not affect the result of the analysis within
the systematic uncertainties assigned.


\section{Systematic Uncertainties}
\label{sec:systematics}

The list of systematic uncertainties on the measurement of $\sin^2\theta_\text{eff}$
and associated corrections to the fit values, as discussed below,
is shown in Table~\ref{table-systematics}.
These uncertainties arise from both theoretical assumptions and detector modeling.

\begin{table}[tbh]
\begin{center}
\caption{
Corrections to the fit values and systematic uncertainties in the measurement of $\sin^2\theta_\text{eff}$.
}
\begin{tabular}{lcc}
\hline\hline
\vspace{0.1cm}
 source & correction & uncertainty  \\
\hline
 PDF                                         &   -- &  $\pm$0.0013 \\
 FSR                                         &   -- &  $\pm$0.0011 \\
 LO model (EWK)                   &  --  & $\pm$0.0002 \\
 LO model (QCD)                   &   +0.0012 &  $\pm$0.0012 \\
 resolution and alignment    &   +0.0007 &  $\pm$0.0013 \\
 efficiency and acceptance  &   --   &  $\pm$0.0003 \\
 background                           &   --   &   $\pm$0.0001 \\
\hline
 total                                        & +0.0019 & $\pm$0.0025 \\
\hline\hline
\end{tabular}
\label{table-systematics}
\end{center}
\end{table}

We follow the PDF4LHC working group's recommendation~\cite{Botje:2011sn} in estimating
uncertainties from the PDFs. We reweight a large MC sample generated with
\textsc{ct10}~\cite{Lai:2010} PDFs to obtain samples equivalent to \textsc{mstw 2008}~\cite{Martin:2009iq}
and \textsc{nnpdf 2.1}~\cite{Ball:2010de} PDFs. We vary the internal degrees of freedom of the PDFs
for all three sets of models. We also use this technique to vary $\alpha_s$, but find its
uncertainties to have negligible effects compared to any of the PDF variations.
We find a change in the value of $\sin^2\theta_\text{eff}$ of $+4.8\times10^{-4}$ and $+3.4\times10^{-4}$
in using the PDFs from \textsc{mstw} and \textsc{nnpdf}, respectively.
The variations in the value of $\sin^2\theta_\text{eff}$ within each MC simulation due to the PDF
uncertainties are $^{+13.0}_{-12.1}\times10^{-4}$, $^{+3.9}_{-4.1}\times10^{-4}$, and
$\pm 7.3\times10^{-4}$ for \textsc{ct10}, \textsc{mstw}, and \textsc{nnpdf}, respectively.
The envelope of the above variations corresponds to the total systematic uncertainty of $\pm$0.0013.

The FSR is modeled with \textsc{Pythia} in the parameterization of the resolution function  $\mathcal{R}(x)$.
As a cross-check, we use four alternative FSR models for generation, with simplified
detector simulation discussed above: \textsc{Pythia}, \textsc{Photos}~\cite{Golonka:2005pn},
and two different modes in \textsc{Horace}~\cite{Calame:2007}.
All three generator programs perform $O(\alpha)$ calculations of FSR and provide similar
results, leading to differences in the fitted values of $\sin^2\theta_\text{eff}$ of about 0.001.
In addition, the \textsc{Horace} generator allows the exact $O(\alpha)$ calculation and
multiple-photon radiation from all charged states, which is the state-of-the-art EWK calculation.
We found that this has a larger effect on the analysis when a wide range of values for the
dimuon invariant mass $m$ is used. However, with the mass range $80<m<100\GeV$,
the differences in the relevant part of the radiative tail become small.
We perform cross-fits of the four generated samples and the four corresponding resolution
functions $\mathcal{R}(x)$, finding differences in the fitted $\sin^2\theta_\text{eff}$ values of at most 0.0011.
The \textsc{Pythia} sample typically results in larger differences from the other generators,
and the assigned systematic uncertainty of $\pm$0.0011 covers these deviations.
The assigned error conservatively covers the FSR uncertainty in the measurement
when the fraction of FSR radiation is reweighted in simulation. This reweighting technique
is based on the comparison of the FSR fractions between data and the \textsc{Pythia} simulation.

Effects from NLO EWK contributions are expected to be small compared to the statistical
precision of our measurement. Calculations with the \textsc{Zfitter}~\cite{Arbuzov:2005ma}
program indicate that the variation of the effective value of $\sin^2\theta_\text{eff}$ for
light quarks is within 0.0002 of the lepton values. It is only the heavier $\cPqb$ quark that
requires corrections of the order 0.001. However, given that only about 2.5\% of the dileptons
are produced in $\cPqb\cPaqb$ interactions, and no angular asymmetry can be measured
because of the dilution factor with this initial state, these corrections have
a negligible effect on our measurements.

Parameterization of the likelihood function models both the initial-state interactions and the PDFs at LO.
The requirement that the dimuon transverse momentum $p_{\sss T}$ be less than 25\GeV suppresses
the contribution of events with hard jet radiation and reduces the effects from NLO processes.
This requirement also ensures that the transformation between the laboratory frame
and the Collins--Soper frame is small, and that our analytical description of acceptance effects
is correct, without any loss of acceptance coverage.
With generated samples, we observe a bias of $-0.0012\pm0.0006$
in the fit value of $\sin^2\theta_\text{eff}$, which is attributed to NLO effects. In this test,
perfect CMS detector conditions are simulated, which removes most of the detector effects
discussed below. To be conservative, we apply a correction of $+0.0012$ and assign
a systematic uncertainty of $\pm0.0012$ to cover all effects associated with the LO model.
For example, we have investigated the dependence of the expected and observed shift
in the fit result as a function of the $p_{\sss T}$ requirement over a wide range of $p_{\sss T}$. The results
are stable within the uncertainty quoted. The distribution of $p_{\sss T}$ exhibits only a small difference
between the data and the MC simulation. We reweight the MC simulation to match the $p_{\sss T}$
distribution to the data and observe that the results of the fit to the reweighted MC events
are consistent with the case without reweighting to within 0.0004.
Treating the correction as an additive or multiplicative factor does not affect
the final result, as long as the observed value of $\sin^2\theta_\text{eff}$ is close to what is expected.

The detector resolution in the muon track reconstruction is affected by contributions
from the silicon tracker alignment. We perform a realistic simulation of the alignment
procedure to model the statistical precision of the track reconstruction. We observe a bias
of $-0.0013$ in the fit result of $\sin^2\theta_\text{eff}$ when the realistic simulation of
alignment is used in place of perfect conditions. We find that while the statistical
precision of the track reconstruction is well modeled by the realistic simulation, the biases
from $\chi^2$-invariant detector deformations~\cite{Chatrchyan:2009sr} may not necessarily
be well reproduced in the MC simulation. We have investigated nine basic distortions in the
tracker reconstruction geometry, which follow from the cylindrical symmetry of the
system~\cite{Chatrchyan:2009sr, Brown:2008ccb}. In each case, the procedure
of the tracker alignment is repeated after the distortion is introduced.
The effects of the remaining distortions on $\sin^2\theta_\text{eff}$ are all smaller
than $0.001$. The typical initial distortions are taken to be 200 $\mu$m,
which is the approximate value of the constraints from the detector survey,
the placement tolerance, and the observed agreement in the alignment procedure.

In the end, constraints on the above distortions in the tracker reconstruction geometry
come from the data. We have observed that distortions affecting the
$\sin^2\theta_\text{eff}$ fit values also introduce a bias in the mass of the dimuon
pair $m$ as a function of $\cos\theta^*$. We observe a linear trend in the bias
of the average value of $m$ as a function of $\cos\theta^*$, with a slope of $-0.072$ GeV,
in the realistic simulation that is twice as large as that observed in data, $-0.039$ GeV,
when both are compared to simulation with the ideal geometry model.
We also observe a bias in the value of $\sin^2\theta_\text{eff}$ that is twice as large when
an additional systematic distortion is introduced in the realistic simulation, resulting in the
slope of the average $m$ value versus $\cos\theta^*$ dependence also becoming twice as large.
From these studies, we assign a correction of $+0.0007$ to the fit value of
$\sin^2\theta_\text{eff}$ due to alignment effects and a systematic uncertainty of
$\pm0.0013$ to cover the range of possible deviations observed.
In order to minimize the uncertainties from the energy scale bias in the track reconstruction,
the shift of the $\cPZ$ mass in the resolution function $\mathcal{R}(x)$ is left free in the fit,
effectively allowing the energy scale to be determined from the fit to the data.
Consistency between the fit value from the data and the expectation from the MC
simulation is found to be within 0.1\GeV.

We find very weak sensitivity to the efficiency parameterization $\mathcal{G}(Y, \hat{s}, \cos\theta^\ast)$
across the acceptance range because the efficiency is symmetric in $\cos\theta^\ast$.
This leads to negligible effects on the odd terms in the angular distribution that are
sensitive to $\sin^2\theta_\text{eff}$.
The sign of $Y$ is defined by the dimuon system direction along the counterclockwise beam
and has no preferred direction. The sign of $\cos\theta^\ast$ is defined by the charge
of the ``forward" lepton. The cylindrical symmetry of CMS, combined with the random
nature of the ``forward" direction, leads to a symmetry in the efficiency function.
This has been verified with a detailed \textsc{Geant4}--based simulation of the
CMS detector, including calibration and alignment effects. Even in the extreme
case of $\mathcal{G}(Y, \hat{s}, \cos\theta^\ast)$ being flat across
the acceptance range, negligible changes in the fit results are observed with simulated
samples. We also allow parameters of the model to be free in the fit to data. We assign
a systematic uncertainty of $0.0003$ due to efficiency and acceptance parameterization,
which is the level of consistency of results from these studies.

The number of background events $n_\text{bkg}$ is fixed to the expected value
and is varied according to its associated uncertainties.
We assign a 50\% uncertainty to the QCD rate, based on studies with wrong-sign
lepton pairs. The relative size of the sum of the EWK background processes
is expected to be reproduced by simulation to a precision of better than $20\%$.
However, in the mass range $80<m<100\GeV$, the fraction of background
is only $0.05\%$, and the fit results are insensitive to the exact treatment of the background.
The measured $\sin^2\theta_\text{eff}$ value remains stable within $0.0001$,
even when the background is removed from the model.

\section{Results and Discussion}
\label{sec:results}

We have presented a likelihood method to analyze the Drell--Yan process at the LHC.
The process is described by the correlated dilepton rapidity, invariant mass, and decay angle
distributions. The quark direction in the elementary parton collisions, which is not directly
accessible in the proton-proton collisions at the LHC, is modeled statistically using correlations
between the observables.
The result of the analysis, which includes systematic uncertainties and corrections from
Table~\ref{table-systematics}, is
\begin{eqnarray}
\sin^2\theta_\text{eff}=0.2287 \pm 0.0020~(\text{stat.})\pm 0.0025~(\text{syst.})
\,.
\nonumber
\label{eq:sin2thetaW_data}
\end{eqnarray}
This measurement of the effective weak mixing angle in the predominantly
$\cPqu\cPaqu, \cPqd\cPaqd\to \gamma^*/\cPZ\to \mu^-\mu^+$ processes in proton-proton collisions
is consistent with measurements in other processes
\cite{Schael:2005ema, Zeller:2001hh, CDF:2005, D0:2008, Abazov:2011ws, Aktas:2005iv},
as expected within the standard model.

The dominant systematic uncertainties in the measurement include modeling of the PDFs,
FSR, effects beyond the leading order in QCD, as well as detector uncertainties primarily due
to tracker alignment. With increased statistics of the Drell--Yan process at the LHC, a further
reduction of the systematic uncertainties will become critical. Understanding the tracker
alignment will certainly improve as the collaboration gains further experience. Therefore,
we expect the limiting uncertainties to come from the Drell--Yan process modeling.

Uncertainties from PDFs will decrease as better constraints on the proton model become
available from the LHC and elsewhere.
In fact, the Drell--Yan process is itself a useful input to the PDF model constraints, and the methods
discussed in this paper can be used to constrain the parameters in the PDF model. However, one
must be careful not to mix information used for PDF constraints from the Drell--Yan process with
measurements using the same events, unless the correlations are properly taken into account.
Uncertainties from the FSR model may be improved as higher-order electroweak calculations are
integrated with the higher-order QCD calculations of the matrix element in the Drell--Yan process,
such as the incorporation of \textsc{Powheg} and \textsc{Horace}.

The LO approximations in the model may be further improved as NLO matrix elements are
employed in the likelihood approach and more variables are integrated into the analysis.
We view the current LO formalism as a conceptual step in developing multivariate matrix-element
approaches to resonance polarization analyses, which can be applied to precision measurements,
as well as potential new resonances that may be discovered at the LHC.
The evolution of this method may also allow several parameters of the electroweak couplings
to be determined simultaneously, such as a measurement of the vector and axial-vector
couplings of the light quarks separately from the lepton couplings.

\section*{Acknowledgements}

We would like to thank Kirill Melnikov and Alessandro Vicini for useful discussions of NLO QCD and EWK effects.
\hyphenation{Bundes-ministerium Forschungs-gemeinschaft Forschungs-zentren} We wish to congratulate our colleagues in the CERN accelerator departments for the excellent performance of the LHC machine. We thank the technical and administrative staff at CERN and other CMS institutes. This work was supported by the Austrian Federal Ministry of Science and Research; the Belgium Fonds de la Recherche Scientifique, and Fonds voor Wetenschappelijk Onderzoek; the Brazilian Funding Agencies (CNPq, CAPES, FAPERJ, and FAPESP); the Bulgarian Ministry of Education and Science; CERN; the Chinese Academy of Sciences, Ministry of Science and Technology, and National Natural Science Foundation of China; the Colombian Funding Agency (COLCIENCIAS); the Croatian Ministry of Science, Education and Sport; the Research Promotion Foundation, Cyprus; the Estonian Academy of Sciences and NICPB; the Academy of Finland, Finnish Ministry of Education and Culture, and Helsinki Institute of Physics; the Institut National de Physique Nucl\'eaire et de Physique des Particules~/~CNRS, and Commissariat \`a l'\'Energie Atomique et aux \'Energies Alternatives~/~CEA, France; the Bundesministerium f\"ur Bildung und Forschung, Deutsche Forschungsgemeinschaft, and Helmholtz-Gemeinschaft Deutscher Forschungszentren, Germany; the General Secretariat for Research and Technology, Greece; the National Scientific Research Foundation, and National Office for Research and Technology, Hungary; the Department of Atomic Energy and the Department of Science and Technology, India; the Institute for Studies in Theoretical Physics and Mathematics, Iran; the Science Foundation, Ireland; the Istituto Nazionale di Fisica Nucleare, Italy; the Korean Ministry of Education, Science and Technology and the World Class University program of NRF, Korea; the Lithuanian Academy of Sciences; the Mexican Funding Agencies (CINVESTAV, CONACYT, SEP, and UASLP-FAI); the Ministry of Science and Innovation, New Zealand; the Pakistan Atomic Energy Commission; the State Commission for Scientific Research, Poland; the Funda\c{c}\~ao para a Ci\^encia e a Tecnologia, Portugal; JINR (Armenia, Belarus, Georgia, Ukraine, Uzbekistan); the Ministry of Science and Technologies of the Russian Federation, the Russian Ministry of Atomic Energy and the Russian Foundation for Basic Research; the Ministry of Science and Technological Development of Serbia; the Ministerio de Ciencia e Innovaci\'on, and Programa Consolider-Ingenio 2010, Spain; the Swiss Funding Agencies (ETH Board, ETH Zurich, PSI, SNF, UniZH, Canton Zurich, and SER); the National Science Council, Taipei; the Scientific and Technical Research Council of Turkey, and Turkish Atomic Energy Authority; the Science and Technology Facilities Council, UK; the US Department of Energy, and the US National Science Foundation.

Individuals have received support from the Marie-Curie programme and the European Research Council (European Union); the Leventis Foundation; the A. P. Sloan Foundation; the Alexander von Humboldt Foundation; the Belgian Federal Science Policy Office; the Fonds pour la Formation \`a la Recherche dans l'Industrie et dans l'Agriculture (FRIA-Belgium); the Agentschap voor Innovatie door Wetenschap en Technologie (IWT-Belgium); and the Council of Science and Industrial Research, India.

\bibliography{auto_generated}   % will be created by the tdr script.

\cleardoublepage \appendix\section{The CMS Collaboration \label{app:collab}}\begin{sloppypar}\hyphenpenalty=5000\widowpenalty=500\clubpenalty=5000\textbf{Yerevan Physics Institute,  Yerevan,  Armenia}\\*[0pt]
S.~Chatrchyan, V.~Khachatryan, A.M.~Sirunyan, A.~Tumasyan
\vskip\cmsinstskip
\textbf{Institut f\"{u}r Hochenergiephysik der OeAW,  Wien,  Austria}\\*[0pt]
W.~Adam, T.~Bergauer, M.~Dragicevic, J.~Er\"{o}, C.~Fabjan, M.~Friedl, R.~Fr\"{u}hwirth, V.M.~Ghete, J.~Hammer\cmsAuthorMark{1}, S.~H\"{a}nsel, M.~Hoch, N.~H\"{o}rmann, J.~Hrubec, M.~Jeitler, W.~Kiesenhofer, M.~Krammer, D.~Liko, I.~Mikulec, M.~Pernicka, B.~Rahbaran, H.~Rohringer, R.~Sch\"{o}fbeck, J.~Strauss, A.~Taurok, F.~Teischinger, C.~Trauner, P.~Wagner, W.~Waltenberger, G.~Walzel, E.~Widl, C.-E.~Wulz
% ...
\vskip\cmsinstskip
% ...
1:~~Also at CERN, European Organization for Nuclear Research, Geneva, Switzerland\\
% ...

\end{sloppypar}
\end{document}